\documentclass[fleqn,12pt,twoside]{article}
\usepackage{espcrc1}
\usepackage{graphicx}
\usepackage{epsfig}
\usepackage[figuresright]{rotating}
\def\nA{nucleon-nucleus\ }
\def\pA{proton-nucleus\ }
\def\AA{nucleus-nucleus\ }
\def\bbox#1{\mbox{\boldmath $#1$}}
\title{Folding model analysis of elastic and inelastic proton
scattering on Sulfur isotopes}
\author{Dao~T.~Khoa\address{Institute for Nuclear Science {\rm \&}
 Technique, VAEC, P.O. Box 5T-160, Nghia Do, Hanoi, Vietnam.}
 \thanks{Corresponding author, E-mail: khoa@mail.vaec.gov.vn},
 Elias Khan\address{Institut de Physique Nucl\'eaire, IN2P3-CNRS, 91406
  Orsay Cedex, France.},
 Gianluca Col\`o\address{Dipartimento di Fisica and INFN,
 Universit\`a degli Studi, Via Celoria 16, 20133 Milano, Italy.}
 and Nguyen Van Giai$^b$}
\begin{document}
\maketitle
\begin{abstract}
The folding formalism for the \nA optical potential and inelastic form factor
is applied to study elastic and inelastic proton scattering on $^{30-40}$S
isotopes. A recently developed realistic density dependent M3Y interaction,
well tested in the folding analysis of \AA elastic and inelastic scattering, is
used as effective NN interaction. The nuclear ground state and transition
densities (for the 2$^+$ excitations in Sulfur isotopes) are obtained in the
Hartree-Fock-BCS and QRPA approaches, respectively. The best fit ratios of
transition moments $M^n_{2^+}/M^p_{2^+}$ for the lowest 2$^+$ states in Sulfur
isotopes are compared to those obtained earlier in the DWBA analysis of the
same data using the same structure model and inelastic form factors obtained
with the JLM effective interaction. Our folding + DWBA analysis has shown quite
a strong isovector mixing in the elastic and inelastic scattering channels for
the neutron rich $^{38,40}$S nuclei. In particular, the relative strength of
the isovector part of the transition potential required by the inelastic
p+$^{38}$S data is significantly stronger than that obtained with the
corresponding QRPA transition density.
\end{abstract}

\bigskip\noindent{\small Keywords: (p,p') reaction, folding model, QRPA
transition densities, DWBA analysis \\ PACS numbers: 24.10.Eq, 24.10.Ht,
24.50.+g, 25.40.Cm, 25.40.Ep, 21.60.Jz}

%\medskip\noindent{\small\sl Accepted for publication in Nuclear Physics A}
\bigskip\centerline{\small\sl Accepted for publication in Nuclear Physics A}

\section{INTRODUCTION}
The folding model has been used for years to calculate the \nA optical
potential (see, e.g., Refs.~\cite{Si75,Br77,Br78,Ri84,Do98}) and inelastic form
factors (see, e.g., Refs.~\cite{Ma78,Ch85}). It can be seen from the basic
folding formulas that this model generates the first-order term of the
microscopic optical potential that is derived from Feshbach's theory of nuclear
reactions \cite{Fe92}. The success of this approach in describing the observed
\nA elastic scattering data for many targets suggests that the first-order term
of the microscopic optical potential is indeed the dominant part of the nucleon
optical potential. In the same way, the inelastic (folded) form factor is also
the most important input for the analysis of inelastic scattering data within
the distorted wave Born approximation (DWBA) or coupled-channel approaches.

The basic inputs for a single-folding calculation of the \nA potential are the
nuclear densities of the target and the effective nucleon-nucleon (NN)
interaction. If one has a well tested, realistic effective NN interaction,
the folding model is a very useful approach to check the target nuclear
densities. A popular choice for the effective NN interaction has been one of
the M3Y interactions which were designed to reproduce the G-matrix elements of
the Reid \cite{Be77} and Paris \cite{An83} NN potentials in an oscillator
basis. Although these density {\sl -independent} M3Y interactions were
originally developed for use in the DWBA analysis of (p,p') reaction, they have
been used much more often in the double-folding calculation of the heavy-ion
interaction potential at low and medium energies \cite{Sa79}.

Intensive studies of the refractive nucleus-nucleus scattering during the last
decade has shown that the simple M3Y-type interaction failed to give a good
description of the data, and this has motivated the inclusion into the original
M3Y interactions of an explicit density dependence
\cite{Kh93,Kh95a,Kh95b,Kh97}, to account for the reduction in the (attractive)
strength of the effective NN interaction that occurs as the density of the
nuclear medium increases \cite{Be71}. With parameter values chosen to reproduce
the observed nuclear matter saturation density and binding energy within the
Hartree-Fock (HF) approach, the new density {\sl -dependent} M3Y interactions
have been carefully tested in the folding analysis of the refractive
nucleus-nucleus elastic scattering \cite{Kh93,Kh95a,Kh95b,Kh97}. In the same HF
scheme, parameters of the density dependence of the M3Y interaction are
directly associated with the nuclear incompressibility $K$, and it has been
shown that $K$ values ranging from 240 to 270 MeV are the most appropriate for
the cold nuclear matter \cite{Kh97} (these values are rather consistent with
what can be extracted from the calculations of the isoscalar giant monopole
resonance within the relativistic mean field framework; if non-relativistic
Skyrme or Gogny forces are employed, slightly lower $K$ values are obtained
\cite{Gi01}). The isospin dependence of this interaction was also tested in a
HF study of the asymmetric nuclear matter and in the calculation of the
heavy-ion potential between nuclei with non-zero isospins \cite{Kh96}. It is
therefore of interest to use the new interaction in the folding calculation of
the \pA potential for the analysis of the elastic and inelastic scattering of
exotic nuclei on proton target.

In the present work, we adapt the most recent version of the double-folding
model \cite{Kh00} for the nucleus-nucleus potential to a single-folding
formalism for the elastic and inelastic \nA potentials, using the density- and
isospin dependent M3Y interaction. By using the \nA elastic and transition
potentials folded with the nuclear densities from a self-consistent microscopic
model \cite{Ela00}, we have performed a detailed folding analysis of the recent
elastic and inelastic $^{30,32}$S+p scattering data measured at $E_{\rm
lab}=53A$ MeV \cite{El01} in GANIL, as well as $^{34,36,38,40}$S+p data
\cite{Al85,Ho90,Ke97,Ma99} measured earlier at $E_{\rm lab}=30, 28, 39$ and
$30\ A$ MeV, respectively. The contributions from isoscalar and isovector parts
of the \pA optical potential and inelastic form factors were treated explicitly
in each case to study the isovector mixing effect in the \pA scattering as one
goes along the isotopic chain, from the proton rich $^{30}$S to the neutron
rich (short lived) $^{38,40}$S isotopes.

\section{THE FOLDING MODEL}
\label{sec2}
\subsection{General formalism}
In the folding model, the \pA potential $V$ is evaluated as a Hartree-Fock-type
potential
\begin{equation}
  V_i=\sum_{j\in A}[<ij|v_D|ij>+<ij|v_{EX}|ji>], \label{e1}
\end{equation}
where $v_D$ and $v_{EX}$ are the direct and exchange parts of the effective NN
interaction between the incident proton $i$ and nucleon $j$ in the target $A$
(in the case of $A$ scattering on proton, $A$ is still treated in our formalism
as `target' as given by the inverse kinematics). The antisymmetrization of the
system is done by taking into account the knock-on exchange effects (the
interchange of nucleons $i$ and $j$). Due to the antisymmetrization, the
potential is, in general, nonlocal in coordinate space. An accurate local
approximation is usually obtained by treating the relative motion locally as a
plane wave \cite{Si75,Br77}, and one can reduce the energy-dependent (central)
\pA potential (\ref{e1}) to the following local form
\begin{equation}
 V(E,\bbox{R})=\int\left[\rho(\bbox{r})v_D(\rho,E,s)+\rho(\bbox{R},\bbox{r})
 v_{EX}(\rho,E,s)j_0(k(E,R)s)\right]d^3r,
\label{e4}
\end{equation}
where $\bbox{s}=\bbox{r}-\bbox{R}$, $\bbox{R}$ is the vector joining the
center-of-mass of the target and the incident proton,
$\rho(\bbox{r},\bbox{r}')$ is the nonlocal (one-body) density matrix (DM) for
the target nucleus with $\rho(\bbox{r})\equiv\rho(\bbox{r},\bbox{r})$ and
$\bbox{k}(E,R)$ is the local momentum of relative motion determined as
\begin{equation}
 k^2(E,R)={{2\mu}\over{\hbar}^2}[E_{\rm c.m.}-V(E,R)-V_C(R)].
\label{e5}
\end{equation}
Here, $\mu$ is the nucleon reduced mass, $V(E,R)$ and $V_C(R)$ are,
respectively, the total nuclear and Coulomb potentials evaluated in the
entrance (elastic) channel. The calculation of the {\sl localized} exchange
part of the \nA potential [second term in Eq.~(\ref{e4})] still contains a
self-consistency problem and involves an explicit integration over the nonlocal
nuclear DM. Normally, the density $\rho(\bbox{r})$ is taken either from a
nuclear structure model or directly from electron scattering data. Therefore,
the calculation of the exchange potential in Eq.~(\ref{e4}) is usually done by
using a realistic approximation for the nonlocal DM \cite{Si75,Ca78}
\begin{eqnarray}
 \rho(\bbox{R},\bbox{R}+\bbox{s}) & \simeq & \rho(\bbox{R}+{\bbox{s}\over 2})
 \hat{j_1}\left(k_F(|\bbox{R}+{\bbox{s}\over 2}|)s\right), \nonumber\\
 {\rm where}\ \ \hat{j_1}(x)& = & 3j_1(x)/x\ =\ 3(\sin x-x\cos x)/x^3.
\label{e6}
\end{eqnarray}
To accelerate the convergence of the DM expansion, Campi and Bouyssy
\cite{Ca78} have suggested to choose the local Fermi momentum $k_F(r)$ in the
following form
\begin{equation}
 k_F(r)=\left\{{5\over{3\rho(r)}}\left[\tau(r)-{1\over 4}{\nabla^2\rho(r)}
 \right]\right\}^{1/2} . \label{e8}
\end{equation}
Assuming this prescription, we choose further the extended Thomas-Fermi
approximation \cite{Ri80,Kh01} for the kinetic energy density $\tau(r)$ in the
evaluation of the local Fermi momentum $k_F(r)$. Note that the local
approximations made in our approach are essentially the same as those adopted
earlier in the folding calculations of \nA optical potential
\cite{Si75,Br77,Br78,Ri84} and inelastic form factor \cite{Ma78,Ch85}.

In this connection, it should be noted that there exists a more sophisticated
version of the single-folding model \cite{Do98} where the nonlocal exchange
potential is treated exactly in the Schr\"odinger equation for the scattering
wave function. In this approach one calculates the nonlocal \nA potential using
the explicit expression for each single-particle wave function $|j>$ taken,
e.g., from the shell model. Therefore, this rigorous approach cannot be used in
a general case, when the target wave function is simply represented by a local
density distribution $\rho(\bbox{r})$.

\subsection{Explicit treatment of the isospin dependence}
Exotic nuclei usually have non-zero isospin and it is necessary to make
explicit the isospin degrees of freedom. The spin-isospin decomposition of the
(central) NN interaction is
\begin{eqnarray}
 v_{D(EX)}(\rho,E,s) & = & v^{D(EX)}_{00}(\rho,E,s)+v^{D(EX)}_{10}(\rho,E,s)
 (\bbox{\sigma.\sigma}') \nonumber\\
 & + & v^{D(EX)}_{01}(\rho,E,s)(\bbox{\tau.\tau}')+
 v^{D(EX)}_{11}(\rho,E,s)(\bbox{\sigma.\sigma}')(\bbox{\tau.\tau}').
\label{e10}
\end{eqnarray}
The contribution from the spin dependent terms ($v_{10}$ and $v_{11}$) in
Eq.~(\ref{e10}) to the central \nA potential (\ref{e4}) is exactly zero for a
spin-saturated target. Even for an odd nucleus, this contribution is at most of
$A^{-1}$ effect \cite{Sa83} and is usually neglected in the folding calculation
of the central \nA potential.

By writing the nuclear densities in Eq.~(\ref{e4}) explicitly in terms of the
proton ($\rho_p$) and neutron ($\rho_n$) densities, one can represent the \pA
potential (\ref{e4}) in terms of isoscalar ($V^{IS}$) and isovector ($V^{IV}$)
parts
\begin{equation}
 V(E,\bbox{R})=V^{IS}(E,\bbox{R})+V^{IV}(E,\bbox{R}).
\label{e13}
\end{equation}
%\pagebreak
 Each term in Eq.~(\ref{e13}) has contributions from both the direct
and exchange potentials
\begin{eqnarray}
 V^{IS}(E,\bbox{R}) & = & \int\{[\rho_p(\bbox{r})+\rho_n(\bbox{r})]
 v^D_{00}(\rho,E,s) \nonumber\\
 & + & [\rho_p(\bbox{R},\bbox{r})+\rho_n(\bbox{R},\bbox{r})]
 v^{EX}_{00}(\rho,E,s)j_0(k(E,R)s)\}d^3r.
\label{e14}
\end{eqnarray}
\begin{eqnarray}
 V^{IV}(E,\bbox{R}) & = & \int\{[\rho_p(\bbox{r})-\rho_n(\bbox{r})]
 v^D_{01}(\rho,E,s) \nonumber\\
 & + & [\rho_p(\bbox{R},\bbox{r})-\rho_n(\bbox{R},\bbox{r})]
 v^{EX}_{01}(\rho,E,s)j_0(k(E,R)s)\}d^3r. \label{e15}
\end{eqnarray}
One can see that the $V^{IV}$ term (microscopic form of the symmetry or Lane
potential) is entirely determined by the difference between the proton and
neutron densities.

By using the local approximation (\ref{e6}) for the nonlocal proton and neutron
($\tau=p,n$) density matrices
\begin{eqnarray}
 \rho_\tau(\bbox{R},\bbox{R}+\bbox{s}) & \simeq &
 \rho_\tau(\bbox{R}+{\bbox{s}\over 2})\hat{j_1}\left(k_F^\tau
 (|\bbox{R}+{\bbox{s}\over 2}|)s\right)\ \equiv\
 f_\tau(\bbox{R}+{\bbox{s}\over 2}), \nonumber\\
{\rm where}\ \ k_F^\tau(r) & = & \left\{\left[3\pi^2\rho_\tau(r)\right]^{2/3}+
 {{5C_S[\nabla\rho_\tau(r)]^2}\over{3\rho^2_\tau(r)}}+
 {{5\nabla^2\rho_\tau(r)}\over{36\rho_\tau(r)}}\right\}^{1/2},
\label{e6a}
\end{eqnarray}
the \pA potential (\ref{e14})--(\ref{e15}) can be obtained as
\begin{eqnarray}
 V^{IS(IV)}(E,\bbox{R}) & = & \int[\rho_p(\bbox{R}+\bbox{s})\pm
 \rho_n(\bbox{R}+\bbox{s})]v^D_{00(01)}(\rho,E,s)d^3s \nonumber\\
 & + & \int\left[f_p(\bbox{R}+{\bbox{s}\over 2})\pm
 f_n(\bbox{R}+{\bbox{s}\over 2})\right]
 v^{EX}_{00(01)}(\rho,E,s)j_0(k(E,R)s)d^3s.
\label{e16}
\end{eqnarray}
In Eq.~(\ref{e6a}), $C_S$ is the strength of the so-called Weizs\"acker term
representing the surface contribution to the kinetic energy density. We have
adopted the commonly accepted value $C_S$=1/36 \cite{Ri80} which ensures a fast
convergence of the density matrix expansion. The local Fermi momentum
$k_F^\tau(r)$ is evaluated using the ground state densities only. The local
approximation (\ref{e6a}) was shown in a recent folding analysis \cite{Kh01} to
be at the level of 1\% accuracy.

\subsection{Effective density dependent NN interaction}
The recently parameterized CDM3Y6 interaction \cite{Kh97}, based on the G-matrix
elements of the Paris NN potential \cite{An83}, is used in the present folding
calculation. The energy- and density dependences are factorized out as
\begin{equation}
 v^{D(EX)}_{00(01)}(E,\rho,s)=g(E)F(\rho)v^{D(EX)}_{00(01)}(s).
\label{g1}
\end{equation}
The explicit density dependence was introduced in Ref.~\cite{Kh97}
\begin{equation}
 F(\rho)=C[1+\alpha\exp(-\beta\rho)-\gamma\rho],
\label{g4}
\end{equation}
with parameters adjusted to reproduce saturation properties of nuclear matter
and to yield a nuclear incompressibility $K=252$ MeV in the HF approximation
\cite{Kh93,Kh97}
\begin{equation}
 C=0.2658,\ \alpha=3.8033,\ \beta=1.4099\ {\rm fm}^3,
 \ \gamma=4.0\ {\rm fm}^3.
\label{g5}
\end{equation}
The `intrinsic' energy dependence of the interaction (to be expected if one
regards the effective NN interaction as representing a reaction- or G-matrix of
the Brueckner type \cite{Je77}) is contained in the factor
 $g(E)\approx 1-0.0026\varepsilon$, where $\varepsilon$ is the bombarding energy
per nucleon (in MeV). The radial strengths of the isoscalar and isovector
components of the central M3Y-Paris interaction \cite{An83} can be obtained
\cite{Kh96} in terms of three Yukawas
\begin{equation}
 v^{D(EX)}_{00(01)}(s)=\sum_{\nu=1}^3 Y^{D(EX)}_{00(01)}(\nu)
 {{\exp(-R_\nu s)}\over{R_\nu s}},
\label{g2}
\end{equation}
where the explicit Yukawa strengths are tabulated in Table~\ref{t0}.
\begin{table}\small
\caption{\small Yukawa strengths of the central (\ref{g2}) and spin-orbit
(\ref{g2s}) components of the M3Y-Paris interaction \cite{An83,Kh96}.}
\label{t0}
\begin{tabular}{|c|c|c|c|c|c|c|c|} \hline
$\nu$ & $R_\nu$ & $Y^{D}_{00}(\nu)$ & $Y^{D}_{01}(\nu)$ & $Y^{EX}_{00}(\nu)$ &
$Y^{EX}_{01}(\nu)$ & $Y^{(0)}_{LS}(\nu)$ & $Y^{(1)}_{LS}(\nu)$ \\
  & (fm$^{-1}$) & (MeV) & (MeV) & (MeV) & (MeV) & (MeV) & (MeV) \\ \hline
 1 & 4.0 & 11061.625 & 313.625 & -1524.25 & -4118.0 & -5101.0 & -1897.0 \\
 2 & 2.5 & -2537.5 & 223.5 & -518.75 & 1054.75 & -337.0 & -632.0 \\
 3 & 0.7072 & 0.0 & 0.0 & -7.8474 & 2.6157 & 0.0 & 0.0 \\ \hline
\end{tabular}
\end{table}

In \nA scattering the most important interaction induced by the nucleon spin is
the spin-orbit coupling which is present in both elastic and inelastic
channels. The spin-orbit potential arises naturally in the folding model if the
effective NN interaction itself has a two-body spin-orbit term
\begin{equation}
 v_{LS}(s)\bbox{L.S}\equiv v_{LS}(s)\  {1\over 4}[(\bbox{r}_i-\bbox{r}_j)
 \times (\bbox{p}_i-\bbox{p}_j)].(\bbox{\sigma}_i+\bbox{\sigma}_j)
\end{equation}
For simplicity, we assume that the spin-orbit part of the CDM3Y6 interaction
has the same density- and energy dependences as the central part (\ref{g1})
\begin{equation}
 v_{LS}(E,\rho,s)=g(E)F(\rho)v_{LS}(s).
\label{g1s}
\end{equation}
The radial strength of the spin-orbit components (with the total isospin T=0
and T=1) of the M3Y-Paris interaction \cite{An83} can also be obtained in terms
of Yukawas
\begin{equation}
 v^{(T)}_{LS}(s)=\sum_{\nu=1}^3 Y^{(T)}_{LS}(\nu)
 {{\exp(-R_\nu s)}\over{R_\nu s}},
\label{g2s}
\end{equation}
with the explicit Yukawa strengths tabulated in Table~\ref{t0}.

\subsection{Multipole decomposition}
For a consistent description of the elastic and inelastic scattering by the
folding potential we need to take into account explicitly the multipole
decomposition of the nuclear density \cite{Sa79,Kh00} that enters the folding
calculation (\ref{e13})-(\ref{e16})
\begin{equation}
 \rho_\tau(\bbox{r})=\sum_{\lambda\mu}<J_AM_A\lambda\mu|J_{A'}M_{A'}>C_\lambda
 \rho_{\lambda}^\tau(r)[i^{\lambda}Y_{\lambda\mu}(\hat{\bbox{r}})]^*,
 \ {\rm where}\ \ \tau=p,n;
\label{m1}
\end{equation}
$J_A$ and $J_{A'}$ are the target spins in the ground state and excited state,
respectively. Usually, the $\lambda=0$ term represents the ground state density
(monopole excitation is a special case and not considered here) and a single
multipole $\lambda\neq 0$ dominates in the transition to an excited state with
spin $J_A'$. In such a case, the corresponding term in the sum (\ref{m1})
represents the nuclear transition density for the excited state. Following
Satchler and Love \cite{Sa79}, we have chosen here a normalization such that
$C_0=\sqrt{4\pi}$ and $C_{\lambda}$=1 for $\lambda\neq 0$. The neutron and
proton transition moments are further determined as
\begin{equation}
M^\tau_\lambda=\int_0^\infty \rho_{\lambda}^\tau(r)r^{\lambda+2}dr. \label{m1a}
\end{equation}
We adopt the same definition of the reduced matrix element as that by Brink
and Satchler \cite{Br93}, and the nuclear transition density is such that the
reduced electric transition rate for a $2^{\lambda}$-pole excitation is
obtained from the proton transition moment as
$B(E\lambda\uparrow)=e^2|M^p_\lambda|^2$.

The corresponding multipole decomposition of the folded potential (\ref{e13})
can then be written as
\begin{equation}
V(E,\bbox{R})=\sum_{\lambda\mu}<J_AM_A\lambda\mu|J_{A'}M_{A'}>C_{\lambda}
 \left[V^{IS}_\lambda(E,R)+V^{IV}_\lambda(E,R)\right]
 [i^\lambda Y_{\lambda\mu}(\hat{\bbox{R}})]^*.
\label{m2}
\end{equation}
The isoscalar and isovector parts of the central folded potential (\ref{m2})
consist of the corresponding direct and exchange components
\begin{equation}
V^{IS(IV)}_\lambda(E,R)=V^{IS(IV)}_D(\lambda,E,R)+V^{IS(IV)}_{EX}(\lambda,E,R).
\label{m2a}
\end{equation}
We note that if one writes the multipole expansion of the density (\ref{m1})
explicitly through the proton and neutron deformation parameters $\beta_\tau$,
one would end up with the multipole components of the folded potential that
depend upon those $\beta$ values. Such a version of the folding model will be
suitable for the analysis of inelastic proton scattering from {\em deformed}
nuclei, as has been studied earlier by Hamilton and Mackintosh in their density
dependent folding model \cite{Ma77}.

\subsection{Direct potential}
Using the folding formulas in momentum space \cite{Sa79} the central direct
potential can be calculated simply with the density dependent M3Y interaction
(\ref{g1}). For the elastic scattering we have $J_{A'}=J_A$ and $\lambda=0$,
and the density components $\rho_0^\tau(r)$ in Eq.~(\ref{m1}) are just the
proton and neutron ground state densities. Denoting the total ground state
density as $\rho_0(r)\equiv\rho^p_0(r)+\rho^n_0(r),$ the direct part of the
central elastic potential (\ref{m2a}) can be obtained in the following form
\begin{equation}
 V^{IS(IV)}_D(\lambda=0,E,R)={{g(E)}\over{2\pi^2}}\int_0^{\infty}
 A^{IS(IV)}_0(q)v^D_{00(01)}(q)j_0(qR) q^2dq,
\label{m3}
\end{equation}
where the Fourier transforms of the direct interaction and ground state density
profile are
\begin{eqnarray}
 v^D_{00(01)}(q) & = & 4\pi\int_0^\infty v^D_{00(01)}(r)j_0(qr)r^2dr,
 \nonumber\\
 A^{IS(IV)}_0(q) & = & 4\pi\int_0^{\infty}\left[\rho^p_0(r)
 \pm \rho^n_0(r)\right]F(\rho_0(r))j_0(qr) r^2dr.
\label{m4}
\end{eqnarray}
In evaluating the inelastic form factor (or transition potential) one needs to
include the medium corrections implied by the use of a density dependent NN
interaction (see, e.g., Refs.~\cite{Ch85,Kh00,Fa88}). In a consistent or {\em
`dynamic'} treatment of the density dependence, the change in the density due
to the excitation, $\rho\to\rho_0+\Delta\rho$, will also change the effective
NN interaction
\begin{equation}
 F(\rho)v_{D(EX)}(\rho,s)\to\left[F(\rho_0)+
 \Delta\rho\ {{\partial F(\rho_0)}\over{\partial\rho_0}}
 \right]v_{D(EX)}(s).
\label{m5}
\end{equation}
This prescription happens to be exact for the excitation of a single phonon of
a $2^{\lambda}$-pole harmonic shape vibration \cite{Bro78}. In this case, the
direct transition potential (\ref{m2a}) for a $2^{\lambda}$-pole excitation of
the target can be written in the following form
\begin{equation}
 V^{IS(IV)}_D(\lambda,E,R)={{g(E)}\over{2\pi^2}}\int_0^{\infty}
 \left[A^{IS(IV)}_\lambda(q)+B^{IS(IV)}_\lambda(q)\right]v^D_{00(01)}(q)
 j_\lambda(qR) q^2dq,
\label{m6}
\end{equation}
where the Fourier transforms of the transition density profiles are
\begin{eqnarray}
 A^{IS(IV)}_\lambda(q) & = & 4\pi\int_0^{\infty}
 \left[\rho^p_\lambda(r)\pm\rho^n_\lambda(r)\right]
 F(\rho_0(r))j_\lambda(qr) r^2dr, \nonumber\\
 {\rm and}\ \ B^{IS(IV)}_\lambda(q) & = & 4\pi\int_0^{\infty}
 \left[\rho^p_0(r)\pm \rho^n_0(r)\right]\rho_\lambda(r)
 {{\partial F(\rho_0(r))}\over{\partial\rho_0(r)}}j_\lambda(qr)r^2dr.
 \label{m7}
\end{eqnarray}
Here, the total transition density is denoted as $\rho_\lambda(r)\equiv
\rho^p_\lambda(r)+\rho^n_\lambda(r)$. It is easy to see that
$B^{IS(IV)}_\lambda(q)$ comes from the {\em `dynamic'} treatment of the density
dependent interaction discussed in Eq.~(\ref{m5}).

\subsection{Exchange potential}
The self-consistent (local) exchange potential has to be calculated by an
iterative procedure \cite{Kh00}, and the exchange part of the elastic potential
(\ref{m2a}) can be evaluated as
\begin{equation}
 V^{IS(IV)}_{EX}(\lambda=0,E,R)=2\pi g(E)\int_0^\infty
 G^{IS(IV)}_0(R,s)v^{EX}_{00(01)}(s)j_0(k(E,R)s)s^2 ds,
 \label{m8}
\end{equation}
\begin{eqnarray}
{\rm where}\ \ G^{IS(IV)}_0(R,s) = \int_{-1}^1\left[f^p_0(y(x),s)
 \pm f^n_0(y(x),s)\right]F(\rho_0(y(x))dx, \nonumber\\
 y(x) = \sqrt{R^2+\displaystyle{s^2\over 4}+Rsx}\ \ {\rm and}\ \
 f^\tau_0(y,s)=\rho^\tau_0(y) \hat{j_1}\left(k_F(y)s\right).
\label{m9}
\end{eqnarray}

After some integral transformation using the expansion formula \cite{Va88} for
spherical harmonics of the mixed argument
$Y_{\lambda\mu}(\widehat{\bbox{R}+\bbox{s}})$, the exchange transition potential
(\ref{m2a}) for a $2^{\lambda}$-pole excitation of the target can be written
as
\begin{eqnarray}
 V^{IS(IV)}_{EX}(\lambda,E,R) & = & 2\pi g(E)\int_0^\infty
 \left[G^{IS(IV)}_\lambda(R,s)+H^{IS(IV)}_\lambda(R,s)\right] \nonumber\\
 & \times & v^{EX}_{00(01)}(s)j_0(k(E,R)s)s^2 ds,
 \label{m10}
\end{eqnarray}
where the exchange kernels are
\begin{eqnarray}
 G^{IS(IV)}_\lambda(R,s) & = & R^\lambda\int_{-1}^1
 \left[f^p_\lambda(y(x),s)\pm f^n_\lambda(y(x),s)\right]
 {{F(\rho_0(y(x))}\over{[y(x)]^\lambda}}dx, \nonumber\\
 H^{IS(IV)}_\lambda(R,s) & = & R^\lambda\int_{-1}^1\left[f^p_0(y(x),s)
 \pm f^n_0(y(x),s)\right]{{\rho_\lambda(y(x))}\over{[y(x)]^\lambda}}
 \left[{{\partial F(\rho_0(y(x)))}\over{\partial\rho_0(y(x))}}\right]dx,
\label{m11}
\end{eqnarray}
with $f^\tau_\lambda(y,s)=\rho^\tau_\lambda(y)\hat{j_1}\left(k_F(y)s\right).$
Note that we have implied the same dynamic treatment of the density dependence
(\ref{m5}) which leads to the $H^{IS(IV)}_\lambda(R,s)$ kernel.

\subsection{Spin-orbit and Coulomb potentials}
In the elastic channel, in addition to the central potential
$V^{IS(IV)}_0(E,R),$ the \nA optical potential has also the spin-orbit part
$V_{LS}(\lambda=0,E,R)(\bbox{l}.\bbox{\sigma})$. In many cases, a
phenomenological Thomas form for the spin-orbit potential suffices for a good
description of the elastic data if the strength is adjusted properly. Within
the folding model, the spin-orbit potential can be evaluated microscopically
using the two-body spin-orbit NN interaction and the nuclear density of the
target. We have chosen the local approximation developed by Brieva and Rook
\cite{Br78} and evaluated $V_{LS}(\lambda=0,E,R)$ using the spin-orbit
component of the CDM3Y6 interaction (\ref{g1s})-(\ref{g2s}), with the isospin
dependence treated explicitly,
\begin{equation}
 V_{LS}(\lambda=0,E,R)=-{{g(E)F(\rho_0(R))}\over 3}\left[\Phi_p(E,R){1\over R}
 {{d\rho^p_0(R)}\over{dR}}+\Phi_n(E,R){1\over R}
 {{d\rho^n_0(R)}\over{dR}}\right],
\label{m12}
\end{equation}
\begin{eqnarray}
 \Phi_p(E,R) & = & \int_0^\infty v^{(1)}_{LS}(s)
  [1+\hat{j_1}(k(E,R)s)]s^4ds, \nonumber\\
 \Phi_n(E,R) & = & {1\over 2}\int_0^\infty
 \{v^{(1)}_{LS}(s)[1+\hat{j_1}(k(E,R)s)]
 +v^{(0)}_{LS}(s)[1-\hat{j_1}(k(E,R)s)]\}s^4ds.
\label{m13}
\end{eqnarray}
In its general formulation, the spin-orbit interaction also contributes to the
inelastic scattering channel and one needs to explicitly calculate the
scattering amplitude in the DWBA formalism, using the spherical tensor
expansion of the spin-orbit interaction and explicit wave function for each
single-particle configuration involved in the excitation \cite{Lo72,Ra97}. Such
a method is not applicable if the target excitation is simply represented by a nuclear
transition density $\rho_\lambda(r)$. Here, we have adopted the same local
approximation  \cite{Br78} as in the elastic case and inserted (\ref{m1}) into
the local density expression. In the first-order approximation, the spin-orbit
transition potential can be obtained as
\begin{equation}
 V_{LS}(\lambda,E,R)=-{g(E)F(\rho_0(R))\over 3}\left[\Phi_p(E,R){1\over R}
 {{d\rho^p_\lambda(R)}\over{dR}}+\Phi_n(E,R){1\over R}
 {{d\rho^n_\lambda(R)}\over{dR}}\right],
\label{m14}
\end{equation}
where $\Phi_\tau(E,R)$ is determined by the same formula (\ref{m13}). Note that
for the natural-parity excitations considered in this work, the central
transition potential gives a dominant contribution to the inelastic cross
section, and the transition spin-orbit potential (\ref{m14}) has only a minor
effect.

The \pA optical potential or inelastic transition potential, calculated by the
folding model, has to be supplemented by a corresponding Coulomb potential. It
is straightforward to use the same folding method to evaluate microscopically
the \pA Coulomb potential, using the (target) charge density matrix
\begin{equation}
 V_C(E,\bbox{R})=\int{e^2\over{|\bbox{r}-\bbox{R}|}}
 \left[\rho_{\rm charge}(\bbox{r})-\rho_{\rm charge}(\bbox{R},\bbox{r})
 j_0(k(E,R)|\bbox{r}-\bbox{R}|)\right]d^3r.
\label{m15}
\end{equation}
One can use the same local approximation (\ref{e6a}) for the nonlocal proton
density matrix and then take into account the finite proton size\cite{Sa79}
to obtain the charge distribution
 $\rho_{\rm charge}(\bbox{r})$ for the calculation of the Coulomb
potential (\ref{m15}). $V_C(E,\bbox{R})$ can then be expanded into a multipole
series analogous to that of Eq.~(\ref{m2}) for the nuclear potential.

In the optical model (OM) analysis of elastic \pA scattering, the Coulomb
potential $V_{C}(\lambda=0,E,R)$ is usually represented by the Coulomb
potential between a point charge and a uniform charge distribution of radius
$R_{C}=r_CA^{1/3}$. This option for the elastic Coulomb potential can be shown
to have about the same strength and shape at the surface as the microscopic
potential obtained from Eq.~(\ref{m15}). For convenience in referring or
comparing with other results, we have chosen this option for the elastic
Coulomb potential in our OM calculation.

In the DWBA analysis of inelastic \pA scattering with a $2^{\lambda}$-pole
excitation of the target, the transition Coulomb potential $V_{C}(\lambda,E,R)$
has been often taken in a model- and energy independent asymptotic form that
can be expressed in terms of the reduced electric transition rate $B(E\lambda)$
(see, e.g., Refs.~\cite{Kh00,Sa83}). Since a correct input for the Coulomb form
factor is substantial in the analysis of the low-lying electric type
excitations, we have used in the present work the microscopic Coulomb form
factor given by Eq.~(\ref{m15}).

\subsection{Complex elastic and inelastic  potentials}
The original M3Y interaction (\ref{g2}) is real, and the formalism presented
above can be used to generate the real parts of the elastic and inelastic \nA
potentials only. These must be supplemented by imaginary potentials which
account for the absorption into other channels that are allowed energetically
but are not considered explicitly. This absorption contributes imaginary parts
to both the diagonal (elastic) and the off-diagonal (inelastic) potentials. It
is common to resort to a hybrid approach by using the folding model to generate
the real part but to use the standard Woods-Saxon (WS) potential for the
imaginary part, and the local \nA optical potential is
\begin{equation}
 U_0(E,R)=V_0(E,R)+iW_0(E,R)+V_{LS}(\lambda=0,E,R)(\bbox{l}.\bbox{\sigma}),
\label{w1}
\end{equation}
where the real and imaginary parts of the (central) optical potential are given
by
\begin{eqnarray}
 V_0(E,R) & = & V_{C}(R)+N_R[V_D(0,E,R)+V_{EX}(0,E,R)], \nonumber \\
 W_0(E,R) & = & -\frac{W_V}{1+\exp\left((R-R_W)/a_W\right)}
 -\frac{4W_S\ \exp((r-R_W)/a_W)}{[1+\exp((r-R_W)/a_W)]^2}.
\label{w2}
\end{eqnarray}
Here $V_{C}(R)$ is the Coulomb potential between a point charge and a uniform
charge distribution of radius $R_{C}=r_CA^{1/3}$, $R_W=r_WA^{1/3}$ and the real
optical potential $V_{D(EX)}(0,E,R)$ is calculated by using Eqs.~(\ref{m3}) and
(\ref{m8}). The choice of the parameters for the WS imaginary potential can be
made using the available systematics for the nucleon optical potential. The
accurate CH89 global systematics \cite{Va91}, based on several thousands data
sets for the \nA elastic scattering at energies ranging from 10 to 65 MeV, is
used in this work to parameterize the imaginary part of the optical potential.
The normalization factor $N_R$ for the real (folded) optical potential is
adjusted in each case to fit the elastic data. In the present folding approach,
$N_R$ is an approximate way to make small adjustments that may be needed to
take into account the higher order contributions to the (real) microscopic
optical potential, the so-called `dynamic polarization potential' (DPP) in the
Feshbach's formalism \cite{Fe92}. It is obvious that value of $N_R$ should
remain close to unity for this procedure to be reasonable.

We stress that a reaction- or G-matrix calculation for a single nucleon
incident on nuclear matter \cite{Je77} can lead to a complex effective NN
interaction (like the JLM effective interaction), where the absorption is
associated with the finite mean free path of nucleons in {\em nuclear matter}.
The imaginary part of such an effective NN interaction (obtained in a standard
local density approximation) does not account in principle for the important
sources of absorption in {\em finite nuclei} like the excitation of surface
modes, nucleon transfer and breakup reactions, etc. These non-elastic processes
contribute to the energy dependent, nonlocal and complex DPP that gives rise to
the imaginary part of the optical potential \cite{Fe92}.

We apply further the same hybrid scheme to the inelastic (transition)
potential, and the transition potential for a $2^\lambda$-pole excitation is
\begin{equation}
 U_{\lambda}(E,R)=V_{\lambda}(E,R)+iW_{\lambda}(E,R)+
 V_{LS}(\lambda,E,R)(\bbox{l}.\bbox{\sigma}),
\label{w3}
\end{equation}
where the real and imaginary parts of the (central) inelastic potential are
given by
\begin{eqnarray}
 V_{\lambda}(E,R) & = & V_C(\lambda,E,R)+V_D(\lambda,E,R)
 +V_{EX}(\lambda,E,R), \nonumber\\
 W_{\lambda}(E,R) & = & -\delta_{\lambda}
 {{\partial W_0(E,R)}\over{\partial R}}.
\label{w4}
\end{eqnarray}
Thus, the real nuclear, Coulomb and spin-orbit transition form factors are
calculated by the folding approach, while a conventional approach of the
collective vibration model is used to obtain the imaginary transition form
factor. In the present work we have assumed $\delta_{\lambda}$ equal to the
charge deformation length, which is determined from the reduced electric
transition rate $B(E\lambda\uparrow)$ \cite{Sa83} as
\begin{equation}
 \delta_\lambda ={{4\pi\sqrt{B(E\lambda\uparrow)/e^2}}\over
 {(\lambda+2)Z<r^{\lambda-1}>_p}},\ \ {\rm where}\ \
 <r^{\lambda-1}>_p={\displaystyle{\int\rho_0^p(r)r^{\lambda+1}dr}\over
 {\displaystyle\int\rho_0^p(r)r^2dr}}.
 \label{w5}
\end{equation}
The obtained charge deformation lengths for the lowest $2^+$ states in Sulfur
isotopes are given in Table~\ref{t1}. We note that the real central and
spin-orbit inelastic form factors in Eq.~(\ref{w4}) were used in the DWBA
calculation as given by the folding model, without a normalization factor. In
the past, one has often used the same normalization factor $N_R$ for both the
elastic and inelastic folded (central) potentials. Such a method is
`consistent' if one considers $N_R$ of the real (folded) optical potential as
scaling factor of the effective NN interaction for a particular target at the
given incident energy. In our case, parameters of the CDM3Y6 interaction have
been fine-tuned, in the HF calculation, to reproduce the saturation properties
of nuclear matter (with a realistic value of the nuclear incompressibility
$K\approx 252$ MeV \cite{Kh97}) and energy dependence of the \nA optical
potential \cite{Kh93,Kh95c}. Thus, the CDM3Y6 interaction should be suitable
for any target (at different energies) and no further scaling of the
interaction is expected (because it would affect the established HF results, on
which the interaction is based). Therefore, one needs to consider the
normalization factor $N_R$ of the real (folded) optical potential in (\ref{w2})
approximately as an effect caused by contribution of the DPP to the (real)
microscopic optical potential only. Given the effective NN interaction
carefully parameterized, it is reasonable to use the folded inelastic form
factor without re-normalization in the (one-channel) DWBA calculation of
inelastic scattering if the higher-order contribution from the DPP to the real
inelastic folded form factor is weak. We will discuss this aspect again in
Sec.~\ref{sec4}

All the OM and DWBA analyses of elastic and inelastic proton scattering were
made using the code ECIS97 written by J. Raynal \cite{Ra97}.

\section{THE NUCLEAR DENSITIES}
\label{sec3}

Different kinds of nuclear densities can be used as input of the present
folding approach. Actually, Eqs.~(\ref{m1}) and (\ref{m1a}) refer to a
macroscopic description (see, e.g., p.579 of Ref.~\cite{Sa83}), and the
expression (\ref{m1}) should be interpreted as a generic deformed density of an
excited spherical nucleus, in which the $\lambda\neq 0$ multipole terms
represent the difference with respect to the ground state density and may be
used as transition densities for inelastic transition to natural parity states
of multipolarity $\lambda$, provided they are properly normalized. On the other
hand, the present folding approach can naturally accommodate, within its
framework, the microscopic nuclear ground state and transition densities (see
also p.657 of Ref.~\cite{Sa83}) and we believe it is worth to study the
predictive power of the model by using these microscopic nuclear structure
inputs. In this section, we briefly describe the structure model used to obtain
the nuclear ground state and transition densities, having in mind the
application of the model to the neutron rich Sulfur isotopes and giving some
details relevant to this case.

We work in the framework of self-consistent calculations based on
Hartree-Fock plus Random Phase Approximation (HF+RPA) with the use of Skyrme
interactions. This nuclear structure model has been used for a long time to
determine the structure properties of the nuclear vibrational states,
especially, the giant resonances. Recently it has been extended to include
pairing correlations by performing quasi-particle RPA (QRPA) calculations on
top of HF+BCS \cite{Ela00,El01} (see also Ref.~\cite{Col01}). In the study of
nuclei along extended isotopic chains the inclusion of pairing is indeed
called for, and its effect is expected to be important in the open shell
isotopes, especially in the case of low-lying excitations.

\subsection{The nuclear ground state densities}

In this work, we perform spherical HF+BCS calculations in coordinate space
using the SGII parameterization \cite{Gi81} of the Skyrme interaction. We
choose a simple prescription for the pairing interaction by assuming a constant
pairing gap given by $\Delta=12/\sqrt{\rm A}$ MeV \cite{Bo69}. The use of a
constant pairing gap produces unrealistic results (states at relatively high
energy acquire quite large occupation probabilities) unless a cutoff is set in
the single-particle space such that states above this cutoff do not feel any
pairing interaction. In the case of the Sulfur isotopes, this cutoff is chosen
such as to include all the subshells of the major shell to which the Fermi
level belongs. The ground state neutron and proton densities are then obtained
as
\begin{equation}\label{eq:bcs}
\rho_0^\tau(r) = \sum_{\alpha}v_{\alpha}^2\varphi_{\alpha}(r),
\end{equation}
where $\varphi_{\alpha}(r)$ and
$v_{\alpha}$ are the radial wave function and BCS amplitude of the
quasi-particle state $|\alpha>$, respectively. The summation on $\alpha$ runs
over neutrons (protons) for $\tau$=n ($\tau$=p).

\subsection{The nuclear transition densities}

We briefly recall the main steps of the QRPA calculations \cite{Ela00}. Based
on the HF+BCS quasi-particle basis, the QRPA equations are solved in the
configuration space, using the standard matrix form. The residual interaction
between quasi-particles is derived from the Skyrme force used in the HF+BCS
calculation, without including the pairing contribution. The continuous part of
the single quasi-particle spectrum is discretized by diagonalizing the HF+BCS
Hamiltonian on a harmonic oscillator basis. The size of the QRPA model space,
i.e., the number of two quasi-particle configurations included, is chosen large
enough to exhaust the appropriate energy-weighted sum rule. Solving the QRPA
equations provides us with the energies $\omega_\kappa$ as well as the wave
functions of the excited states $|\kappa>$ of a given multipolarity $\lambda$.
These wave functions are normally given in terms of the well-known amplitudes
$X^\kappa_{\alpha\beta}$ and $Y^\kappa_{\alpha\beta}$, with $\alpha\beta$
denoting the involved two quasi-particle configurations. The nuclear transition
to the excited state $|\kappa>$ can be characterized by the corresponding
(local) proton or neutron transition densities
$\Delta\rho_\kappa^\tau(\bbox{r})$ defined as
\begin{equation}
 \Delta\rho_\kappa^\tau(\bbox{r})\equiv <\kappa\ |\sum_i
 \delta(\bbox{r}-\bbox{r}_i)|\ {\rm g.s.}>, \label{eq:tran}
\end{equation}
where the summation on $i$ runs over neutrons (protons) for $\tau$=n ($\tau$=p).

\begin{table}\small
\caption{\small The excitation energy $\omega$, reduced electric transition
probability $B(E2\uparrow)$ and ratio of quadrupole transition moments
$M=(M^n_{2^+}/M^p_{2^+})/(N/Z)$ for the lowest 2$^+$ states in Sulfur isotopes,
as given by QRPA (see Sec.~\ref{sec3}). The `experimental' $M_{\rm exp}$ ratio
is obtained by scaling the proton part of the QRPA transition density to
reproduce the experimental $B(E2)_{\rm exp}$ value and adjusting the neutron
part to the best DWBA fit to the inelastic scattering data. The
`phenomenological' $M_{\rm Phenom}$ ratio was obtained in
Refs.~\cite{El01,Ma99} based on standard collective model form factor. The
experimental values for $\omega_{\rm exp}$ and $B(E2)_{\rm exp}$ are taken
from Refs.~\cite{Ra87,Sc96,Gl97}. The charge deformation lengths $\delta_2$
are determined from the corresponding $B(E2)_{\rm exp}$ values using
Eq.~(\ref{w5}).} \label{t1}
\begin{tabular}{|c|c|c|c|c|c|c|c|c|} \hline
 Nucleus & $\omega_{\rm QRPA}$ & $\omega_{\rm exp}$ & $B(E2)_{\rm QRPA}$ &
 $B(E2)_{\rm exp}$ & $\delta_2$ & $M_{\rm QRPA}$ & $M_{\rm exp}$ &
 $M_{\rm Phenom}$  \\
  & (MeV) & (MeV) & ($e^2$fm$^4$) & ($e^2$fm$^4$) & (fm) &  &  & \\ \hline
 $^{30}$S & 2.79 & 2.21 & 327 & 320$\pm 40$ & 1.19 & 1.05 & 1.05 &
 $0.93\pm 0.20$ \\
 $^{32}$S & 2.94 & 2.23 & 294 & 300$\pm 13$ & 1.14 & 0.96 & 0.96 &
 $0.95\pm 0.11$ \\
 $^{34}$S & 2.65 & 2.13 & 256 & 212$\pm 12$ & 0.95 & 0.94 & 1.04 &
 $0.91\pm 0.11$ \\
 $^{36}$S & 3.46 & 3.29 & 241 & 96$\pm 26$ & 0.63 & 0.64 & 0.90 &
 $1.13\pm 0.27$ \\
 $^{38}$S & 2.19 & 1.30 & 325 & 235$\pm 30$ & 0.98 & 0.98 & 1.44 &
 $1.50\pm 0.30$ \\
 $^{40}$S & 1.54 & 0.89 & 431 & 334$\pm 36$ & 1.17 & 1.03 & 1.17 &
 $1.25\pm 0.25$ \\ \hline
\end{tabular}
\end{table}

It is evident from this definition (see also p.657 of Ref.~\cite{Sa83}) that
these are the appropriate quantities to be folded with the NN interaction to
obtain the inelastic \pA form factors as described in Sec.~\ref{sec2}. The
explicit radial form of the QRPA transition density, for the transition to an
excited state $|\kappa>$ of the multipolarity $\lambda$, is
\begin{equation}\label{eq:tr}
\Delta\rho_{\kappa}^\tau(r)\equiv\rho_{\lambda}^\tau(r)=\sum_{\alpha\geq\beta}
\varphi_{\alpha}(r)\varphi^{*}_{\beta}(r)<\beta||Y_{\lambda}||\alpha>
\lbrace{X^\kappa_{\alpha\beta}- Y^\kappa_{\alpha\beta}}\rbrace
\lbrace{u_{\alpha}v_{\beta}+(-1)^\lambda v_{\alpha} u_{\beta}}\rbrace~.
\end{equation}

The HF+BCS ground state (\ref{eq:bcs}) and QRPA transition densities
(\ref{eq:tr}) of the considered Sulfur isotopes are those obtained recently by
the Orsay group \cite{El01} (see, e.g., the radial shape of these densities in
Figs.~10 and 11 of Ref.~\cite{El01}). In particular, the QRPA charge
ground-state and transition densities for $^{32,34}$S have been shown to
reproduce reasonably well the experimental charge densities of these isotopes
(measured directly by electron scattering). It should be noted that these
transition densities have been previously used to obtain the inelastic S+p form
factors using the JLM effective interaction \cite{El01}. Therefore, the
comparison of the results obtained for the Sulfur isotopic chain will lend
crucial information on the validity of reaction models (see further discussion
in Sec.~\ref{sec4}).

Among the integral properties of the QRPA transition densities, the most
important ones are the reduced transition probabilities $B(E\lambda\uparrow)$
and the ratio of the transition moments $M^n_{\lambda}/M^p_{\lambda}$, as
determined by Eq.~(\ref{m1a}). In a simple collective model \cite{Bo69,Be83},
the oscillation of the homogeneous neutron-proton fluid is so-called purely
isoscalar if $M^n_{\lambda}/M^p_{\lambda}=N/Z$. Consequently, a significant
deviation of $M^n_\lambda/M^p_\lambda$ from $N/Z$ would indicate the degree of
what we call the {\em isovector mixing} in the considered quadrupole
excitations. We will discuss this as well as other structure effects in more
details in the next Section.

\section{RESULTS AND DISCUSSION} \label{sec4}
\subsection{Elastic scattering}

\begin{figure}[htb]\vspace*{-2cm}
\begin{minipage}[t]{80mm}
\hspace*{-1.8cm}\vspace*{-4cm} \mbox{\epsfig{file=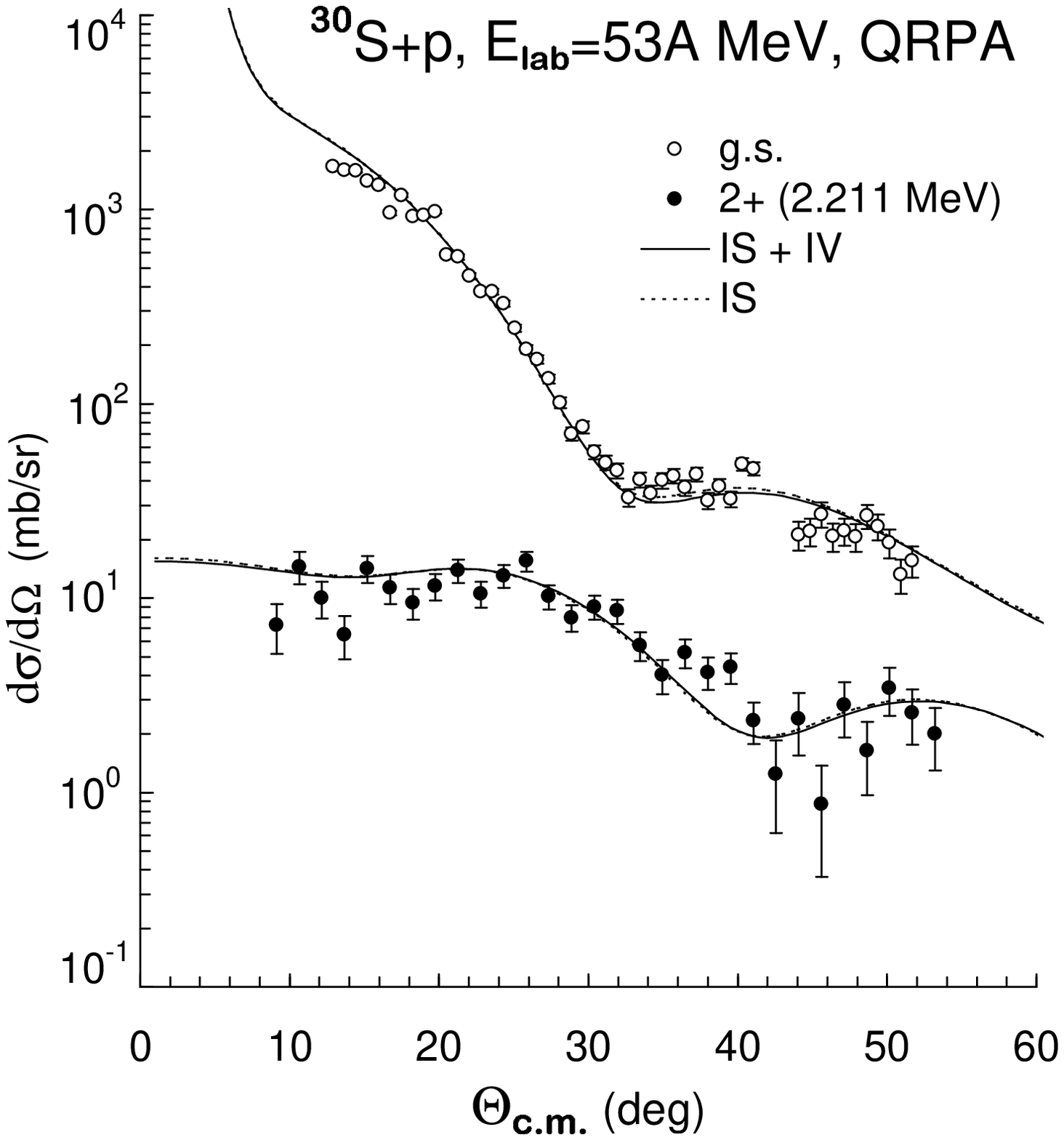,height=13.5cm}}
\end{minipage}
\hspace{\fill}
\begin{minipage}[t]{75mm}
\hspace*{-1.2cm}\vspace*{-4cm} \mbox{\epsfig{file=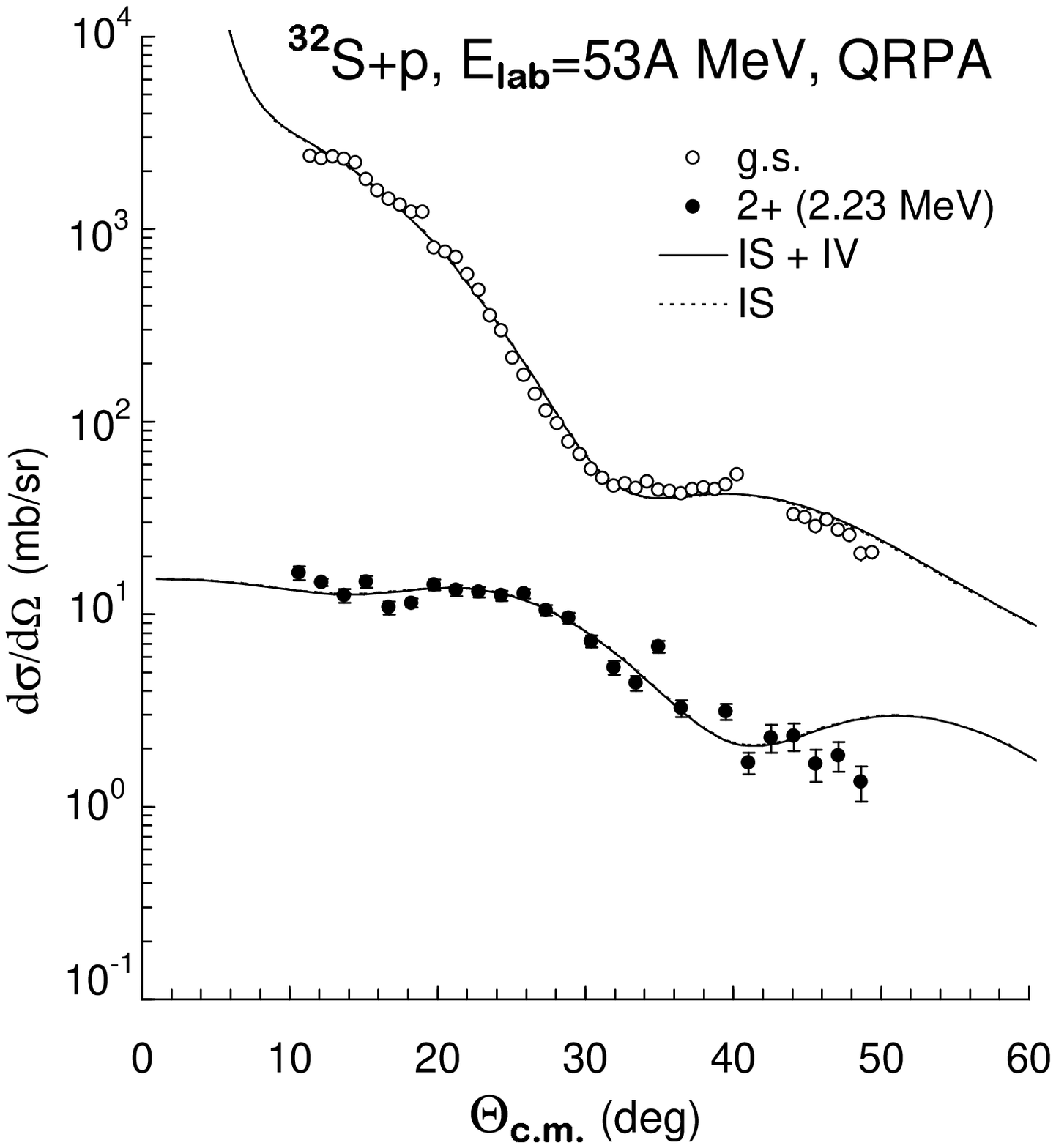,height=13.5cm}}
\end{minipage} \vspace*{-4cm}
\caption{\small Elastic and inelastic $^{30,32}$S+p scattering data at $E/A=53$
MeV \cite{El01} in comparison with the DWBA cross sections given by the elastic
and inelastic potentials folded with the HF+BCS ground-state and QRPA
transition densities, respectively. The cross sections given by the isoscalar
potentials alone are plotted as dotted curves.} \label{f1}
\end{figure}

\begin{figure}[htb]\vspace*{-2cm}
\hspace*{2cm}\vspace*{-4cm} \mbox{\epsfig{file=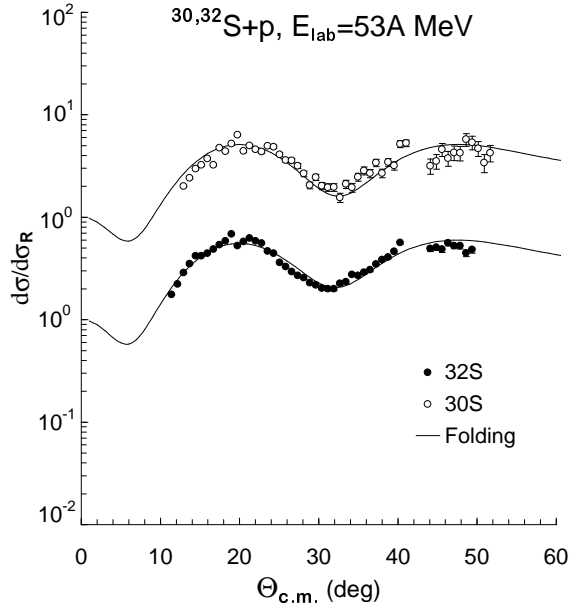,height=14cm}}
\caption{\small Elastic $^{30,32}$S+p scattering data at $E/A=53$ MeV
\cite{El01} and results of the present OM calculation plotted in ratio to
corresponding Rutherford cross section.} \label{f1c}
\end{figure}

\begin{figure}[htb]\vspace*{-2cm}
\begin{minipage}[t]{80mm}
\hspace*{-1.8cm}\vspace*{-4cm} \mbox{\epsfig{file=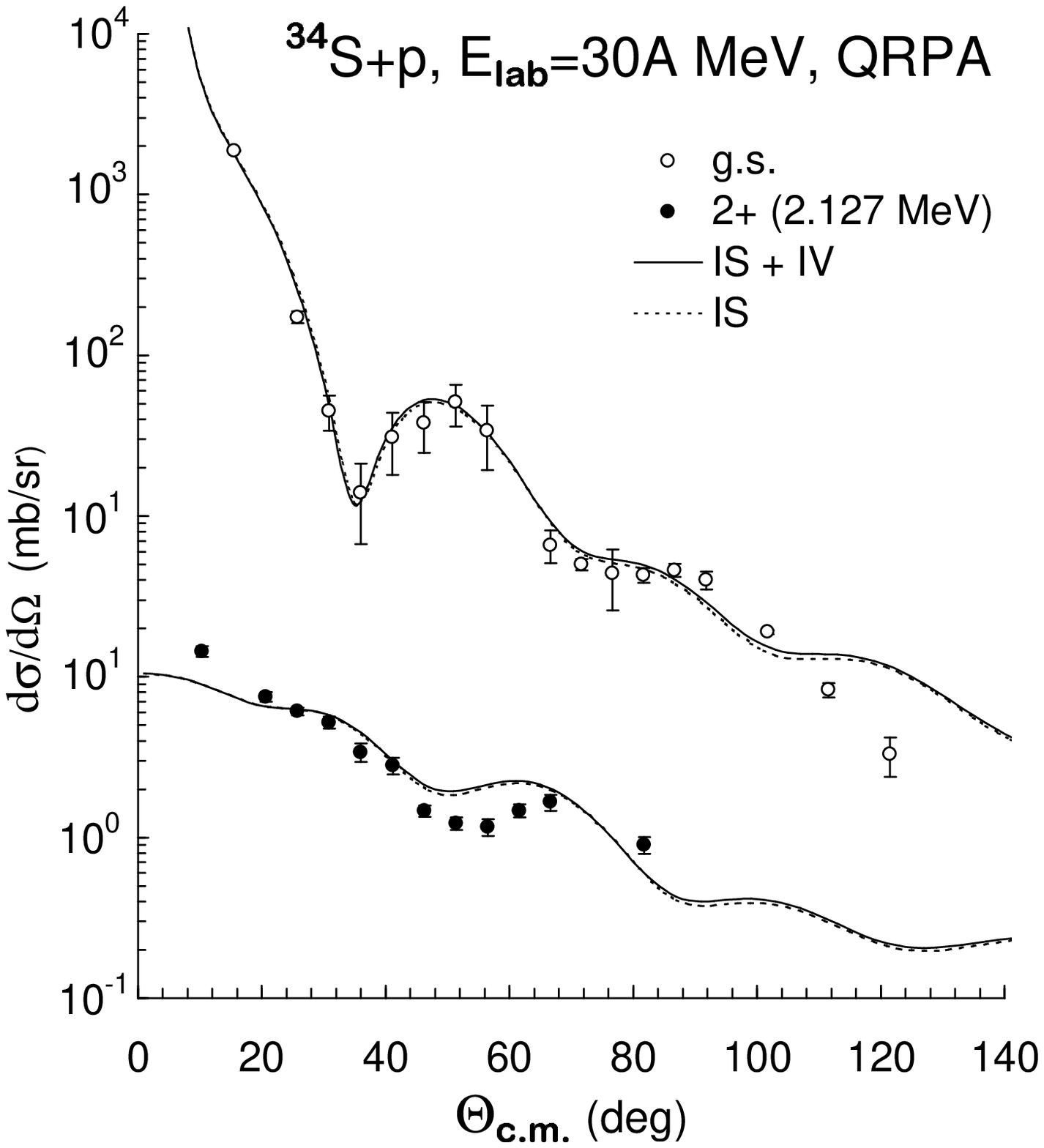,height=13.5cm}}
\end{minipage}
\hspace{\fill}
\begin{minipage}[t]{75mm}
\hspace*{-1.2cm}\vspace*{-4cm} \mbox{\epsfig{file=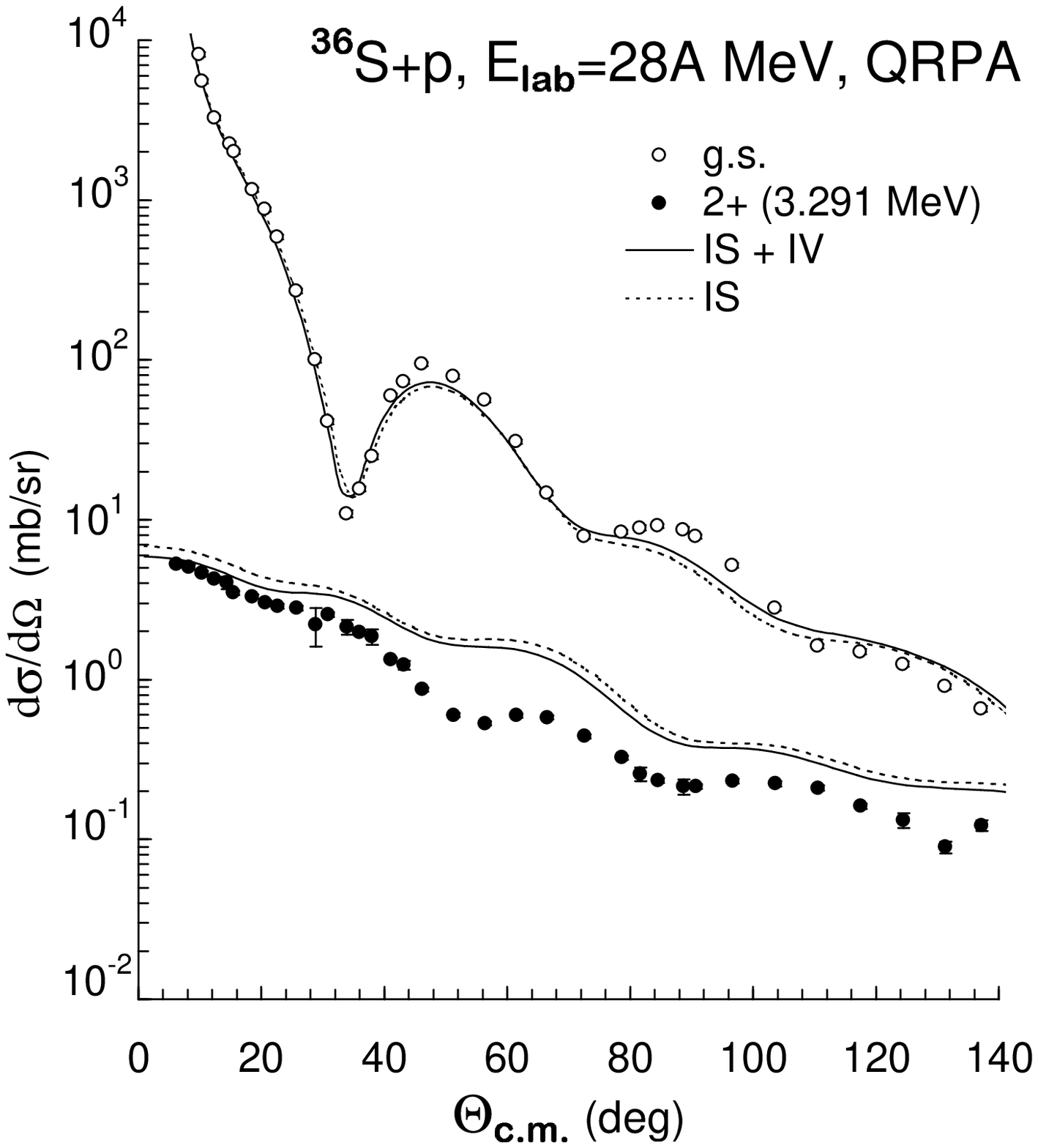,height=13.5cm}}
\end{minipage}
\vspace*{-4cm} \caption{\small The same as Fig.~\ref{f1} but for elastic and
inelastic $^{34}$S+p and $^{36}$S+p data at $E/A=30$ \cite{Al85} and 28 MeV
\cite{Ho90}, respectively.} \label{f2}
\end{figure}

For any \nA system, the accurate measurement of the elastic scattering provides
essential information for the determination of the \nA optical potential, which
is further used to generate the distorted waves for the calculation of
inelastic scattering amplitude in the DWBA formalism. The considered elastic
$^{30-40}$S+p scattering data at different energies have been analyzed in the
present work with the optical potential obtained from Eqs.~(\ref{w1}) and
(\ref{w2}). The real optical potential was obtained using the CDM3Y6
interaction and the ground-state densities given by the HF+BCS calculation,
while the WS imaginary potential was parameterized using the CH89 global
systematics \cite{Va91}. The best-fit optical model parameters are given in
Table~\ref{t2}. From the results plotted in upper parts of
Figs.~\ref{f1}-\ref{f3} one can see that our folding potential reproduces the
elastic data quite well. Although the CH89 parameters are mainly based on the
\nA elastic data for medium-mass targets, they seem to be able to provide a
reasonable estimate for the Sulfur isotopes as well. In the $^{30}$S+p and
$^{40}$S+p cases, the CH89 parameters of the WS imaginary part of the optical
potential are very appropriate and only the renormalization factor $N_R$ of the
real folded potential was slightly adjusted by the OM search to fit the elastic
data. For the $^{32-38}$S isotopes, the strengths $W_V, W_S$ were also adjusted
by the OM fit and the calculated elastic cross sections agree very well with
the measurement. The obtained OM parameters (Table~\ref{t2}) have been used
further in all the DWBA calculations of inelastic scattering cross sections.

We note that the recently measured elastic $^{30,32}$S+p data at $E/A=53$ MeV
\cite{El01} are quite accurate and have more data points compared to other
cases. To show the reliability of the semi-microscopic optical potential
obtained in the present approach, we have further plotted in Fig.~\ref{f1c} the
elastic $^{30,32}$S+p data and the calculated cross section in ratio to
Rutherford cross section (because the actual quality of the OM fit then becomes
more transparent). One can see that the oscillating structure of the elastic
cross section becomes more pronounced if plotted as ratio to Rutherford, and
our OM results give indeed a very good fit to the measured data.

In recent years, various experimental studies of reaction cross sections have
been done to probe the nuclear sizes, especially the existence of halo or
neutron skin in unstable nuclei \cite{Oz01}. It is, therefore, of interest to
present the calculated total reaction cross sections $\sigma_R$ for the
considered cases (see Table~\ref{t2}). From the relative difference
$\Delta\sigma_R$ between the total reaction cross section given by our folding
analysis and that given by the mass- and energy dependent systematics for
stable isotopes (see Eq.~(1) in Ref.~\cite{Ca96}), one finds that our results
agree with the systematics within about 10\%. $\Delta\sigma_R$ is only 5\% for
$^{32}$S (a stable, $N=Z$ isotope), and we do not see any enhancement in
$\sigma_R$ for the proton-rich $^{30}$S isotope that might indicate a proton
halo in this nucleus. To provide crucial data on this point, measurements at
large center-of-mass angles are called for. Note that the total reaction cross
section is mainly determined by the imaginary part of the optical potential,
and the large angle data, if available, would reduce the uncertainty in the
absorption strength and $\sigma_R$ values can be extracted more precisely.
Among the Sulfur isotopes under study $\Delta\sigma_R$ is largest for $^{40}$S
(about 18\%) which is clearly due to the neutron skin in this neutron-rich
(unstable) isotope. The neutron (proton) r.m.s. radii calculated within HF+BCS
for this nucleus are 3.44 fm (3.24 fm), as reported in Table 2 of
Ref.~\cite{El01}.

\begin{table}\small
\caption{\small OM parameters [see Eqs.~(\ref{w1}) and (\ref{w2})] used in the
folding analysis of the elastic proton scattering on Sulfur isotopes. The real
optical potentials were folded with the CDM3Y6 interaction and ground-state
densities given by the HF+BCS calculation. Starting parameters of the WS
imaginary optical potential and Coulomb radius were taken from the global
systematics CH89 \cite{Va91}. Parameters which were fixed during the OM search
are given in boldface.} \label{t2}
\begin{tabular}{|c|c|c|c|c|c|c|c|c|c|} \hline
 Target & $E/A$ & $N_R$ & $W_V$ & $W_S$ & $r_W$ & $a_W$ & $r_C$ &
 $\sigma_R$ & $\Delta\sigma_R$ \\
  & (MeV) &  & (MeV) & (MeV) & (fm) & (fm) & (fm) & (mb) & (\%) \\ \hline
 $^{30}$S & 53 & 0.940 & {\bf 5.347} & {\bf 3.723} & {\bf 1.195} &
 {\bf 0.690}& {\bf 1.275} & 669.4 & 12 \\
 $^{32}$S & 53 & 0.901 & 4.037 & 3.480 & {\bf 1.198} & {\bf 0.690}&
 {\bf 1.275} & 643.3 & 5 \\
 $^{34}$S & 30 & 0.930 & 3.949 & 6.315 & {\bf 1.200} & {\bf 0.690}&
 {\bf 1.274} & 937.5 & 12 \\
 $^{36}$S & 28 & 0.947 & 1.074 & 6.968 & {\bf 1.203} & {\bf 0.690}&
 {\bf 1.273} & 945.7 & 11 \\
 $^{38}$S & 39 & 0.935 & 4.315 & 4.809 & {\bf 1.205} & {\bf 0.690}&
 {\bf 1.273} & 894.5 & 10 \\
 $^{40}$S & 30 & 0.990 & {\bf 2.571} & {\bf 7.926} & {\bf 1.207} &
 {\bf 0.690}& {\bf 1.272} & 1079 & 18 \\ \hline
\end{tabular}
\\ $\Delta\sigma_R$ is the relative difference between the total reaction cross
section given by our folding analysis and that given by the mass- and energy
dependent systematics for stable isotopes \cite{Ca96}.
\end{table}

In an OM study of the elastic scattering induced by a neutron-rich nucleus, the
isospin dependence of the optical potential should become more significant.
While this effect is negligible in $^{30,32}$S+p cases, it becomes more and
more sizable in $^{36,38,40}$S+p cases. From the prediction of the HF+BCS
calculation for the neutron rich $^{36,38,40}$S isotopes (see, e.g., Fig.~10 in
Ref.~\cite{El01}), the extra neutrons do not form the halo-like structure and
they distribute evenly both in the interior and at the surface. As a result,
the ground state neutron density is larger than the proton density at all
radii. This difference determines the isovector part of the optical potential
$V^{IV}_0(E,R)$, i.e., the microscopic estimation of the Lane potential
\cite{La62}. Our results show that the maximal difference in the elastic
scattering cross section, caused by the isovector potential $V^{IV}_0(E,R)$, is
ranging from about 15\% in $^{36}$S+p case up to about 30\% in $^{40}$S+p case
(see upper parts of Figs.~\ref{f2} and \ref{f3}). Although the difference found
is still within the experimental errors of the measured elastic cross section,
this result indicates that the isovector component of the \pA optical potential
might be probed in accurate measurements of the elastic scattering induced by
neutron rich beams. Such experiments could be an alternative study to (p,n)
charge exchange reaction (where the isovector component of the \nA optical
potential can be separately tested \cite{Co98,Vi01}).

\subsection{Inelastic scattering}
The present folding analysis of inelastic scattering data has been performed
within the standard DWBA using the transition form factors defined in
Eqs.~(\ref{w3}) and (\ref{w4}). Since the effective NN interaction has been
fine-tuned in the HF studies of symmetric \cite{Kh97} and asymmetric
\cite{Kh96} nuclear matter, it is expected to be a quite reliable input for the
folding calculation and the test of the QRPA transition densities in the
folding + DWBA analysis. One can see in Table~\ref{t1} that the proton parts of
the QRPA transition densities for the lowest 2$^+$ states in $^{30,32}$S give
$B(E2)$ values very close to the experimental ones. The corresponding neutron
transition densities were found to have about the same radial shape (see
Fig.~11 in Ref.~\cite{El01}) and they give, therefore,
$M^n_{2^+}/M^p_{2^+}\approx N/Z$. From the DWBA results plotted in
Fig.~\ref{f1} one can see that the inelastic form factor calculated from the
QRPA transition densities describes very well the inelastic $^{30,32}$S+p
scattering data measured recently at $E_{\rm lab}=53A$ MeV \cite{El01}. This
confirms the microscopic structure of the lowest 2$^+$ states in $^{30,32}$S
predicted by the QRPA. Since the difference between the proton and neutron
transition densities is very small for these two nuclei (especially for
$^{32}$S) the isovector component of the folded form factor is negligible and
the DWBA cross section is determined mostly by the isoscalar form factor. In
other words, the lowest quadrupole excitation in $^{30,32}$S is predominantly
isoscalar, with the best-fit moment ratio
 $M=(M^n_{2^+}/M^p_{2^+})/(N/Z)\approx 1$.

We note that, given the accurate QRPA transition density (which has a proton
part very close to the experimental charge transition density \cite{El01}) and
well tested CDM3Y6 interaction \cite{Kh97}, the new proton scattering data
measured for the ($N=Z$) stable $^{32}$S isotope turn out to be a very good
test case for our folding approach. The fact that, without any scaling of the
folded form factor, the DWBA cross section agrees perfectly with the inelastic
$^{32}$S+p data confirms the reliability of the present folding + DWBA approach
in probing the structure of the lowest 2$^+$ states in Sulfur isotopes. This
result also shows that the inelastic form factor folded with the CDM3Y6
interaction (which has been fine-tuned in the HF studies) can enter the DWBA
calculation without a re-normalization factor, i.e., $N_R=1$ for the inelastic
folded potential. As a result, the higher-order contribution from the DPP to
the {\em real} inelastic form factor seems to be weaker than that to the real
optical potential.

The DWBA folding results are compared with the $^{34,36}$S+p scattering data in
Fig.~\ref{f2}. With the predicted $B(E2)$ value about 20\% larger than the
experimental data for $^{34}$S (see Table~\ref{t1}), the agreement between the
DWBA cross section with the inelastic data is still reasonably good for the
$^{34}$S+p system. In $^{36}$S, the $B(E2)$ value given by the QRPA is roughly
2.5 times larger than the experimental one, and the DWBA cross section for the
2$^+$ excitation in $^{36}$S consequently overestimates the data over the whole
measured angular region, especially at large scattering angles. The damping of
the lowest 2$^+$ excitation in $^{36}$S, leading to a transition strength
$B(E2)_{\rm exp}$ about two times smaller than that observed for the lowest
2$^+$ states in neighboring Sulfur isotopes, cannot be explained within the
current QRPA model. The neutron transition strength predicted by QRPA turns out
to be quite weak and gives a moment ratio $M=(M^n_{2^+}/M^p_{2^+})/(N/Z)\approx
0.64$. Several structure models together with the DWBA analysis of the same
data using the form factor folded with the JLM effective interaction
\cite{El01,Ma99} have also failed to describe this nucleus. Thus, $^{36}$S
nucleus still remains a puzzle to theoreticians, and the structure models
should start being able to reproduce its $B(E2)$ value. We will see below that
some disagreement with the measured angular distribution remains after
renormalizing the QRPA (proton) transition density to reproduce the
experimental $B(E2)_{\rm exp}$ value.

\begin{figure}[htb]\vspace*{-2cm}
\begin{minipage}[t]{80mm}
\hspace*{-1.8cm}\vspace*{-4cm} \mbox{\epsfig{file=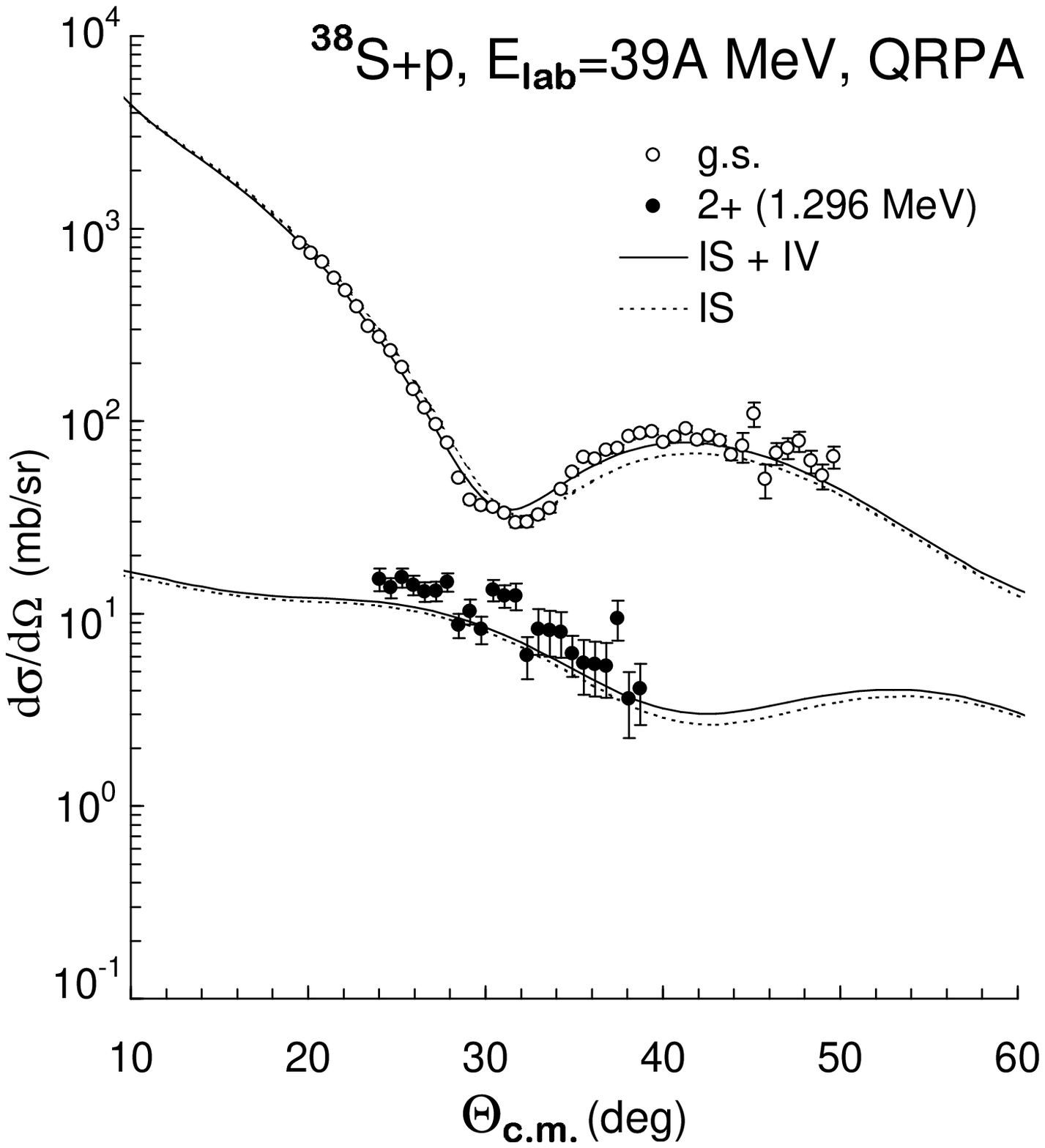,height=13.5cm}}
\end{minipage}
\hspace{\fill}
\begin{minipage}[t]{75mm}
\hspace*{-1.2cm}\vspace*{-4cm} \mbox{\epsfig{file=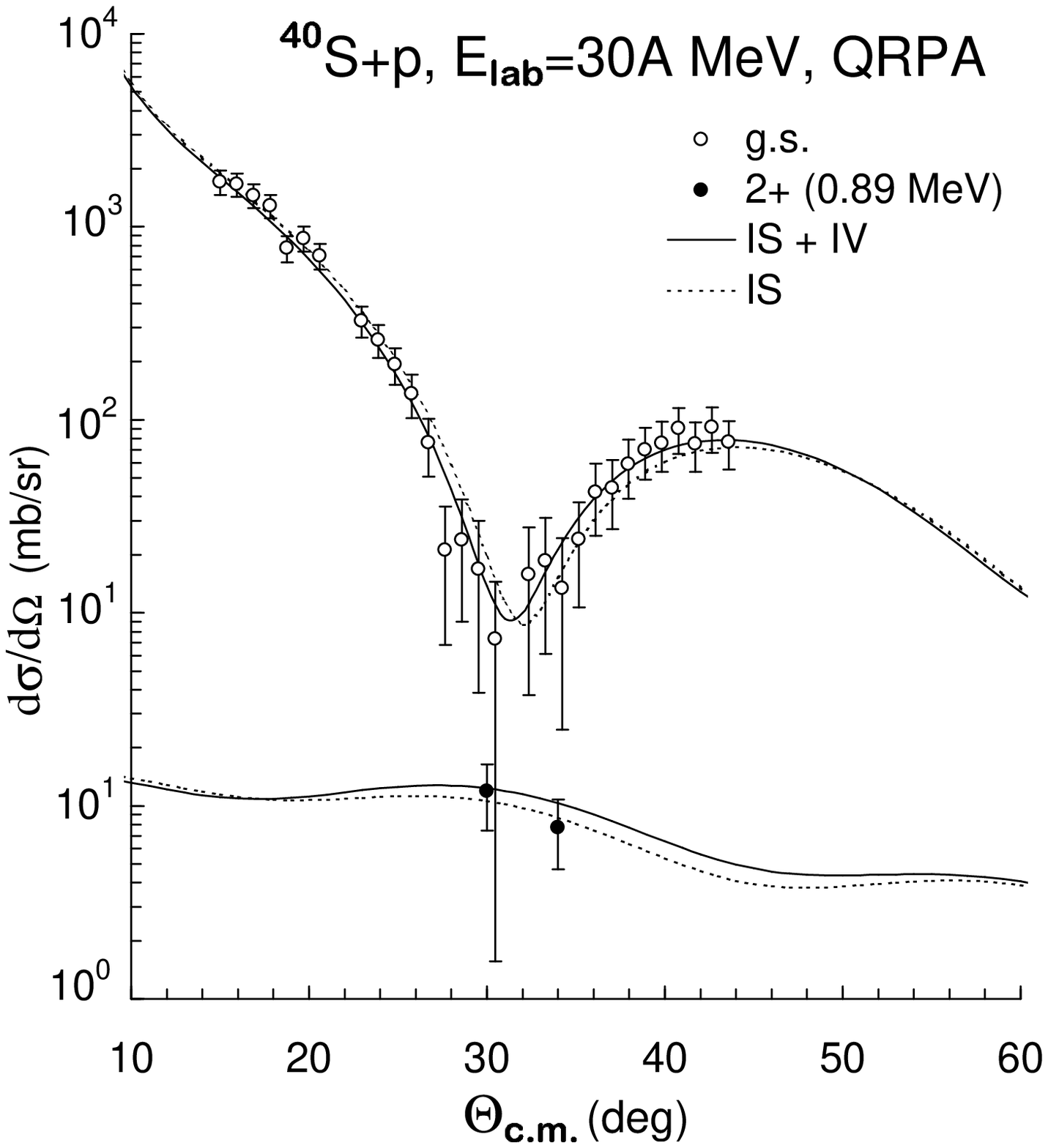,height=13.5cm}}
\end{minipage}
\vspace*{-4cm} \caption{\small The same as Fig.~\ref{f1} but for elastic and
inelastic $^{38}$S+p and $^{40}$S+p data at $E/A=39$ \cite{Ke97} and 30 MeV
\cite{Ma99}, respectively.} \label{f3}
\end{figure}

The DWBA results for $^{38,40}$S+p systems are compared with the scattering
data in Fig.~\ref{f3}. Even though the predicted $B(E2)$ value is about 40\%
larger than the experimental one in $^{38}$S (see Table~\ref{t1}),
the DWBA cross section underestimates the inelastic $^{38}$S+p data by about
30\%. This indicates that the QRPA neutron transition strength is somewhat
weaker than that required by the inelastic data. The predicted moment ratio
$M^n_{2^+}/M^p_{2^+}$ for $^{38}$S  is close to $N/Z$ (the same as in
$^{32}$S) and it could also indicate a lack of neutron transition
strength in the QRPA results. For $^{40}$S, the predicted $B(E2)$
is about 30\% larger than the experimental value and $M^n_{2^+}/M^p_{2^+}\simeq
N/Z$ (see Table~\ref{t1}). However, it is difficult to discuss about the proton
and neutron transition strengths of the $2^+$ state in $^{40}$S based only on
two data points \cite{Ma99} (with rather large experimental errors).

\begin{figure}[htb]\vspace*{-2cm}
\begin{minipage}[t]{80mm}
\hspace*{-1.8cm}\vspace*{-4cm} \mbox{\epsfig{file=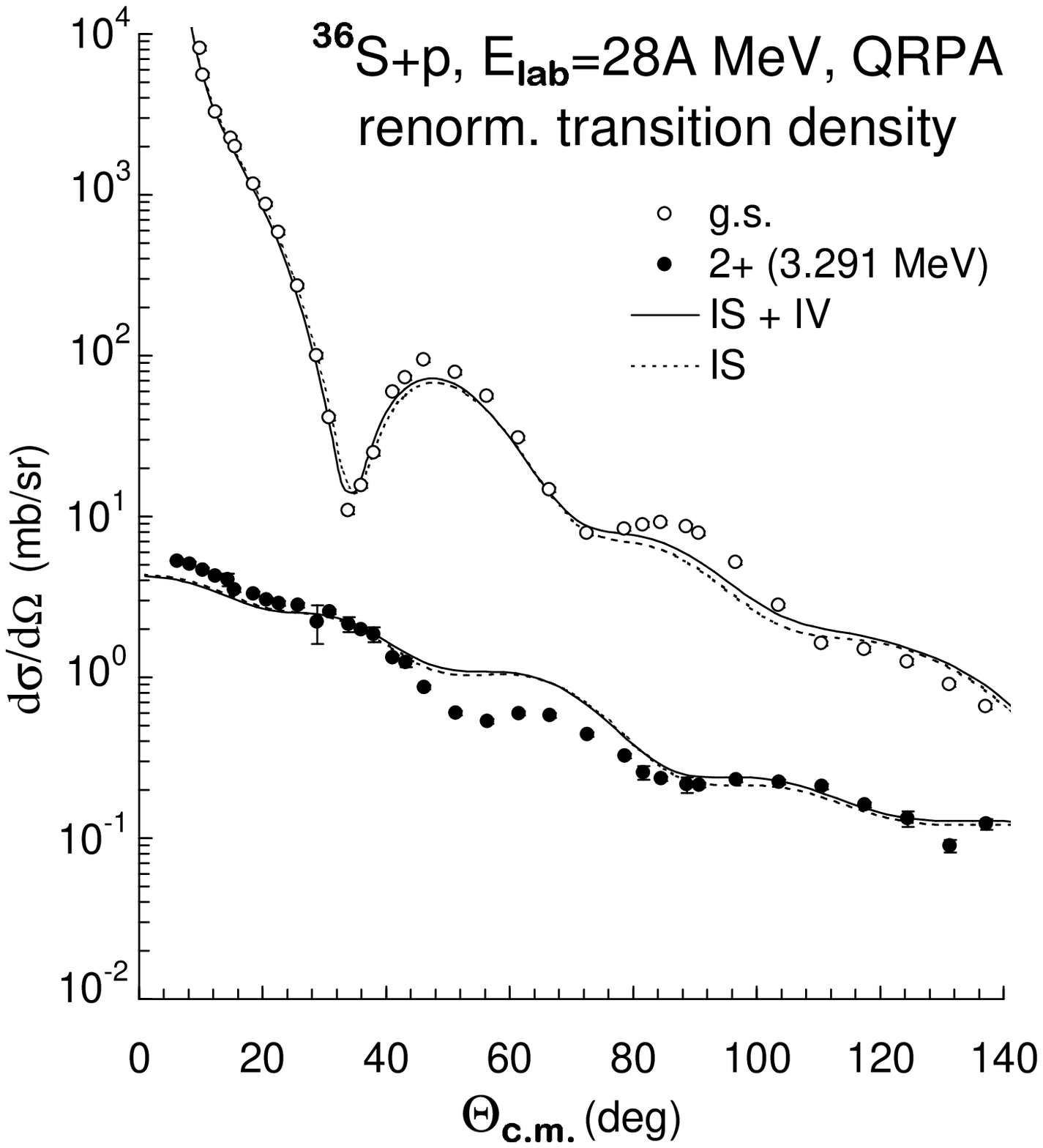,height=13.5cm}}
\end{minipage}
\hspace{\fill}
\begin{minipage}[t]{75mm}
\hspace*{-1.2cm}\vspace*{-4cm} \mbox{\epsfig{file=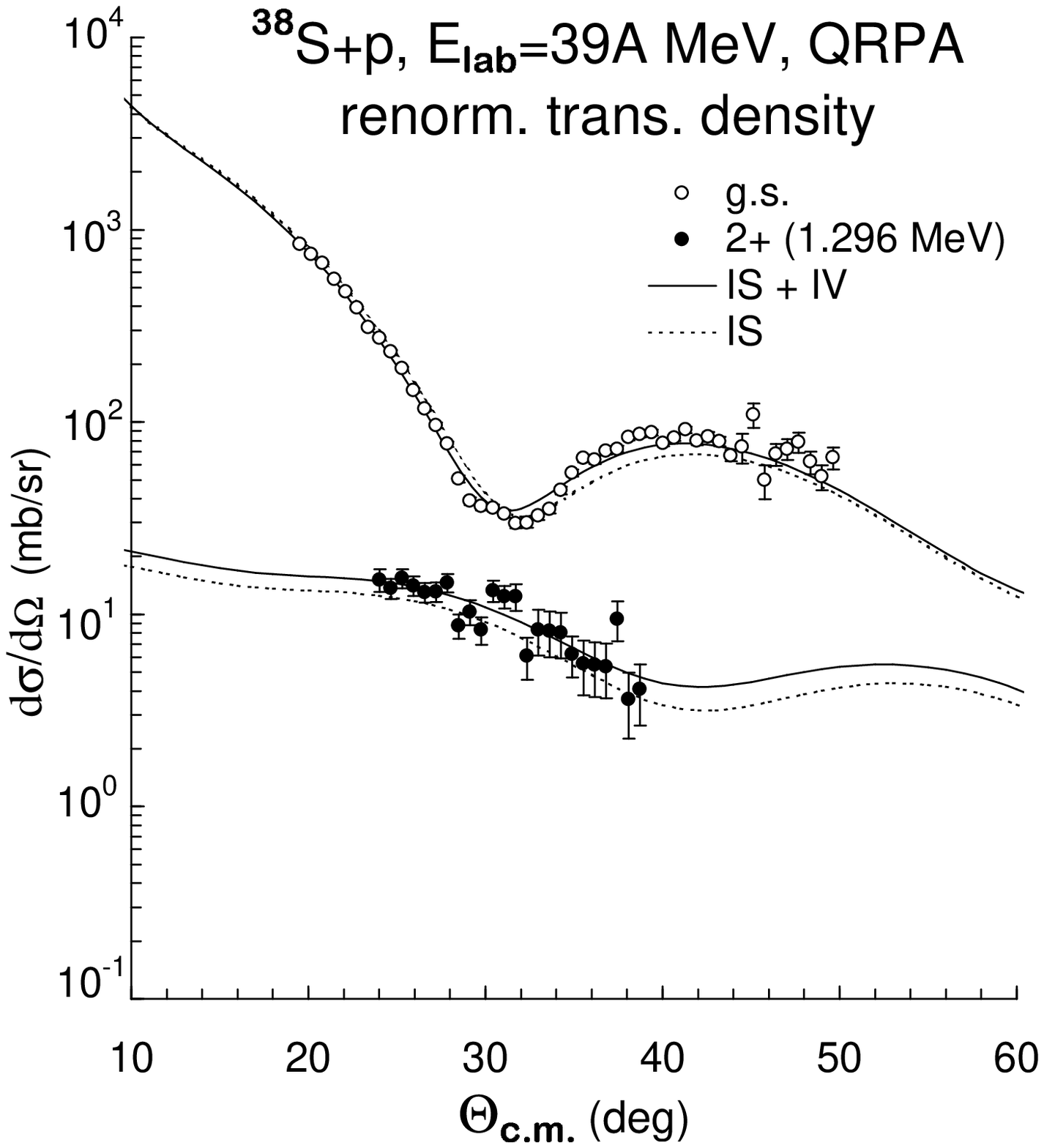,height=13.5cm}}
\end{minipage}
\vspace*{-4cm} \caption{\small The same as Fig.~\ref{f1} but for elastic and
inelastic $^{36}$S+p and $^{38}$S+p data at $E/A=28$ \cite{Ho90} and 39 MeV
\cite{Ke97}, respectively. Inelastic (transition) form factors folded with
 {\em renormalized} QRPA transition densities were used in the DWBA
 calculation (see details in Text).} \label{f4}
\end{figure}

Clearly, one can discuss quantitatively the transition strengths
only when the theoretical model gives good description of both the
$B(E2)_{\rm exp}$ value and the measured inelastic angular distribution. For
that purpose, we have further made the same folding + DWBA analysis of
$^{34-40}$S+p systems using an inelastic form factor obtained from {\em
renormalized} QRPA transition densities, as done earlier by Khan {\sl et
al.} \cite{El01}. Namely, the proton part of the QRPA transition density is
scaled to give the experimental $B(E2)_{\rm exp}$ value and the neutron part
is adjusted by the best DWBA fit to the inelastic scattering data. The
moment ratio obtained in this way is denoted as $M_{\rm exp}$ in
Table~\ref{t1}. Note that for $^{30,32}$S isotopes, the QRPA transition
densities give good description to both the $B(E2)_{\rm exp}$ value and the
measured inelastic scattering data, so that $M_{\rm exp}=M_{\rm QRPA}$.
Among other cases, the improved agreement between the DWBA cross section and
the data is clearly seen in Fig.~\ref{f4} for the $^{36,38}$S+p systems.
One can notice that, even after the QRPA transition density is
renormalized, there remains some disagreement with the inelastic $^{36}$S+p
data at medium angles. The scattering at medium and large angles is known to
probe the form factor at sub-surface distance, and this effect might
indicate a deficiency in the radial shape of the QRPA transition density
for $^{36}$S which cannot be changed by the renormalization
procedure. The damping of the $2^+$ excitation in $^{36}$S can be
judged by the scaling factors of the QRPA transition density: after the
proton transition density is scaled down by a factor of 0.63 to give
$B(E2)=B(E2)_{\rm exp}\approx 96\ e^2$fm$^4$, the neutron transition density
needs to be renormalized by a factor of 0.88 to fit the inelastic data.
Although the obtained moment ratio ($M_{\rm exp}\approx 0.90$) is larger
than that predicted by QRPA for $^{36}$S, it still shows some
damping of the neutron transition strength.

\begin{figure}[htb]\vspace*{-2cm}
\begin{minipage}[t]{80mm}
\hspace*{-1.8cm}\vspace*{-4cm} \mbox{\epsfig{file=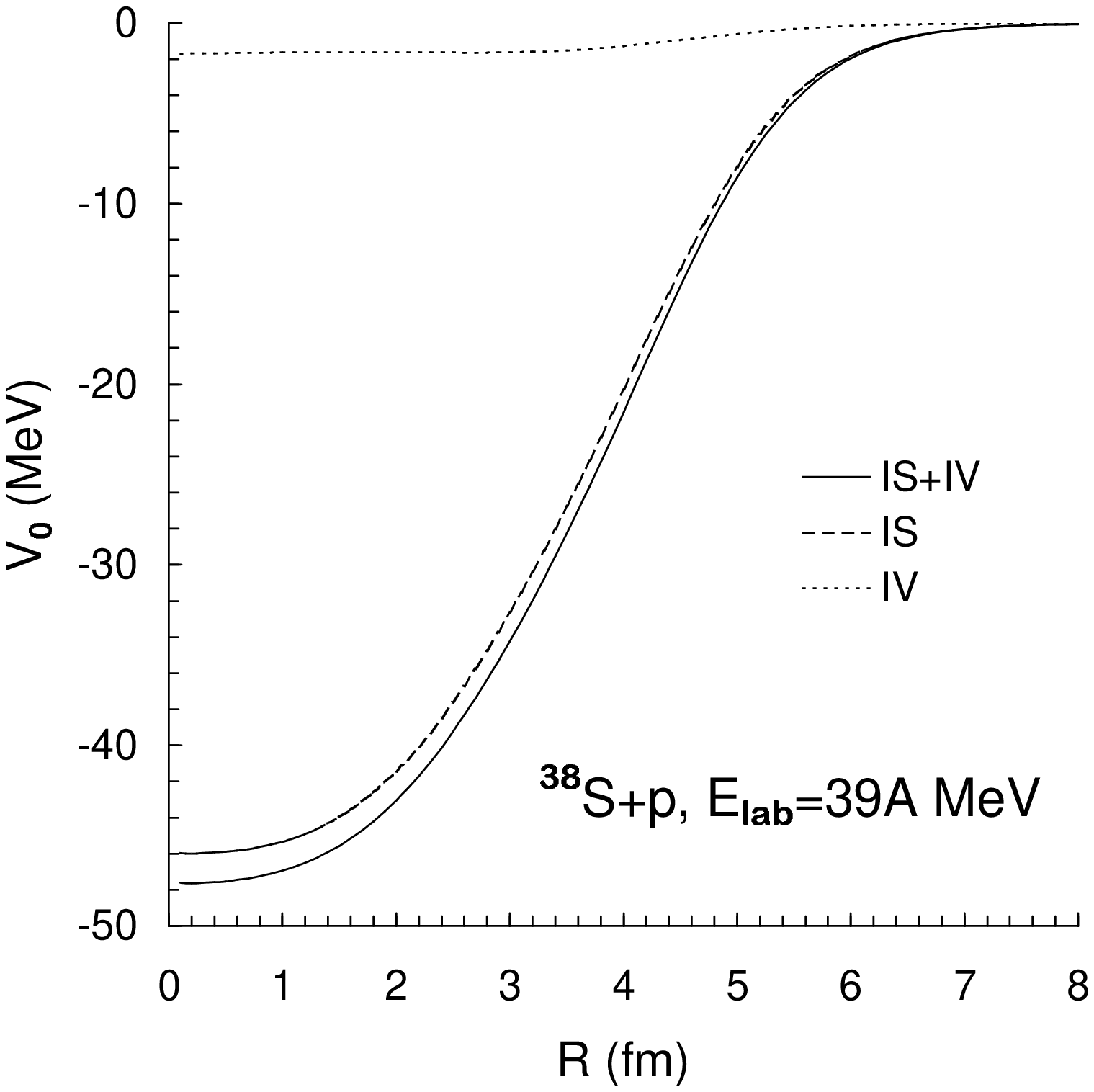,height=13.5cm}}
\end{minipage}
\hspace{\fill}
\begin{minipage}[t]{75mm}
\hspace*{-1.2cm}\vspace*{-4cm} \mbox{\epsfig{file=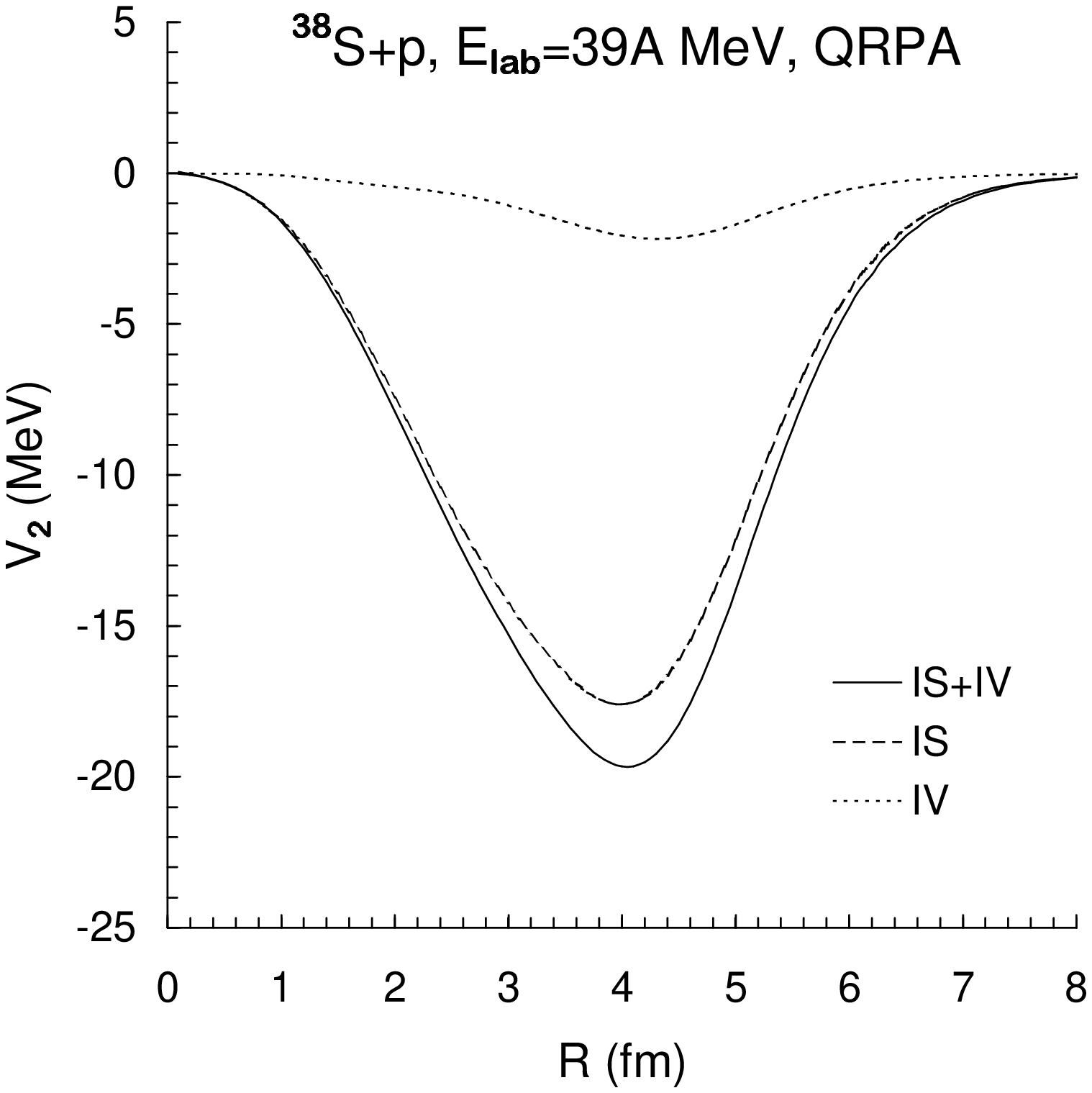,height=13.5cm}}
\end{minipage}
\vspace*{-4cm} \caption{\small $^{38}$S+p (real) optical potential (left panel)
and inelastic form factor for the 2$^+$ excitation (right panel) folded with
the HF+BCS ground-state and {\em renormalized} QRPA transition densities.
Isoscalar and isovector components of the potentials are plotted as dashed and
dotted curves, respectively.} \label{f5}
\end{figure}

The results obtained for the $^{38}$S+p system are quite interesting. After the
proton transition density was scaled by a factor of 0.85 to give the
experimental $B(E2)_{\rm exp}$ value, the neutron transition density needed to
be renormalized by a factor of 1.25 for the best agreement of DWBA cross
section with the inelastic data. As a result, the difference between the proton
and neutron transition densities becomes larger and enhances the isovector
contribution to the DWBA cross section. The 2$^+$ cross section given by the
isovector part of the $^{38}$S+p form factor is about 25-30\% of the total
2$^+$ cross section. Such a significant contribution is due to a strong neutron
transition strength ($M^n_{2^+}$) compared with the proton one ($M^p_{2^+}$)
which gives the ratio $M_{\rm exp}=1.44$. The isovector mixing is slightly
weaker in the elastic channel, with the cross section given by the isovector
part of the $^{38}$S+p optical potential of about 20\% of the total elastic
cross section. In Fig.~\ref{f5}, one can see that the difference
($\rho_p-\rho_n$) between the proton and neutron transition densities leads to
an enhancement of the isovector part of the folded 2$^+$ form factor [see also
Eqs.~(\ref{m6}) and (\ref{m10})] at distances around $R=4$ fm, where
$V^{IV}_{2^+}$ amounts up to 15\% of the total form factor strength. Since the
inelastic cross section is directly proportional to the square of the form
factor, the 2$^+$ cross section given by the isovector part of the folded form
factor alone (see Fig.~\ref{f4}) becomes around 25-30\% of the
total 2$^+$ cross section. For the elastic scattering the difference
($\rho_p-\rho_n$) between the proton and neutron ground-state densities is
smaller, and the isovector part of the folded optical potential or the Lane
potential is about 7-9\% of the total potential strength at radii near $R=4$ fm
(see left panel of Fig.~\ref{f5}), which leads to an isovector mixing of about
20\% in the elastic $^{38}$S+p cross section.

For $^{40}$S+p system, the isovector mixing becomes stronger in the elastic
channel, with the cross section given by the isovector part of the $^{40}$S+p
folded optical potential of about 30\% of the total elastic cross section.
However, the situation in the $2^+$ inelastic channel is still uncertain
because the best-fit ratio $M_{\rm exp}=1.17$ is based on two data
points only. One clearly needs more experimental information in order to have a
reliable estimate for the transition moment ratio in $^{40}$S.

The comparison of the results using the same structure input but with two
different reaction models (JLM approach and the present folding model) is of
special interest, because of the extensive (p,p') data available along the
Sulfur isotopic chain, including elastic and inelastic angular distributions.
Although the JLM approach gives, in general, good agreement with the data, it
appears to have some difficulties to reproduce both the first minimum and the
elastic distribution at large angles (see Figs.~13, 14, 15 of Ref.~\cite{El01},
Fig.~9 of Ref.~\cite{Ma99} and Fig.~7 of Ref.~\cite{Sc00}). It is interesting
to note that this feature does not happen with the present folding
calculations. For the considered $2^+$ excitations, the DWBA cross sections
given by the inelastic form factors folded with the JLM interaction seem to
overestimate the inelastic data in most cases which results in the smaller
transition moment ratios compared to our study. First, one should note the
radically different approaches of these two models. In the JLM approach, one
uses the local density approximation (LDA) to obtain the local optical
potential from that calculated for {\em infinite} nuclear matter, while in the
folding model both the optical potential and inelastic form factor are
calculated in the Hartree-Fock manner (with antisymmetrization treated
properly) using a fine-tuned density dependent CDM3Y6 interaction that is based
on the G-matrix elements of the Paris NN potential for {\em finite} nuclei.
Therefore, the mentioned problem might well be due to the validity of the LDA
and suggests further investigations to improve the JLM parameters. Another clue
is that a better description of elastic data by the folding potential might
also be due to the fact that the imaginary optical potential is based on CH89
global systematics \cite{Va91}, with its strength slightly adjusted to fit the
elastic data. In this sense, the imaginary WS (elastic and transition)
potentials obtained in our approach could serve as a good reference to improve
the microscopic description of the imaginary part of the optical potential and
the transition form factor for the considered cases. As discussed above, the
$^{36}$S case remains a common problem for both reaction models and the
deficiency should be due to the structure model. Finally we note that the
best-fit transition moment ratios deduced from our folding + DWBA analysis
turned out to be quite close to those deduced earlier (see $M_{\rm exp}$ and
$M_{\rm Phenom}$ values in Table~\ref{t1}) from the DWBA analysis of
essentially the same data \cite{El01,Ma99} using standard collective model form
factor. This is very satisfactory in the sense that the microscopic analysis of
the properties of stable and exotic nuclei, which is the final goal of this
kind of approach, is independent from the reaction model, provided this
reaction model is accurate enough.

\section{CONCLUSION}
A consistent single folding formalism for the \pA optical potential and
inelastic form factor, using a realistic density dependent CDM3Y6 interaction
based on the G-matrix elements of the Paris NN potential and the microscopic
ground-state and transition densities given by the HF+BCS and QRPA
calculations, respectively, has been applied to study elastic and inelastic
scattering of $^{30-40}$S isotopes on proton target. The contribution of the
isovector part of the folded optical potential and inelastic form factor to the
calculated elastic and inelastic cross section has been considered explicitly
in all cases.

For the proton rich $^{32}$S nucleus, the results of the OM and DWBA analyses
reproduce very well the measured elastic and inelastic scattering data and do
not indicate the existence of a proton halo in this nucleus. Our results
have shown quite a strong isovector mixing in the elastic and inelastic
$^{38,40}$S+p scattering. In particular, the best fit strength of the isovector
part of the form factor in the inelastic 2$^+$ channel for $^{38}$S is
significantly stronger than that predicted by the QRPA. The detailed folding +
DWBA analysis of the considered data has closely reproduced the systematics
found earlier (based on standard collective model form factors)
\cite{El01,Ma99} for the neutron-proton ratio of the $2^+$ transition moments
in neutron rich Sulfur isotopes.

In conclusion, the present folding + DWBA approach has been proven to be
accurate and reliable in extracting the neutron and proton transition strengths
$M^{n(p)}_{\lambda}$ of the nuclear excitation induced by the inelastic proton
scattering. Moreover, our approach is very efficient in testing the microscopic
nuclear densities and determining the isospin character of the
considered excitation (in structure of the excited state as well as in the
inelastic angular distribution).

These results confirm again that proton scattering induced by the neutron rich
beams remains a very effective tool in nuclear structure study. Such
measurements would allow us not only to test the ingredients of the microscopic
structure models (like QRPA) but also the isospin dependence of the \pA optical
potential and inelastic form factor (prototypes of the Lane potential) and,
consequently, the isospin dependence of the in-medium NN interaction (which is
a key to specify the equation of state for asymmetric nuclear matter
\cite{Kh96}).

\section*{ACKNOWLEDGEMENTS}
The present research was supported, in part, by Natural Science Council of
Vietnam and Vietnam Atomic Energy Commission (VAEC).

\end{document}